\documentclass[a4paper,12pt]{article}

\usepackage{amssymb}
\usepackage{amsmath}
\usepackage{bm}
\usepackage[dvips]{epsfig}
\usepackage{graphicx}

\begin{document}

\title{Topological integrability, classical and quantum chaos, and 
the theory of dynamical systems in the physics of condensed matter.}

\author{A.Ya. Maltsev$^{1}$, S.P. Novikov$^{1,2}$}

\date{
\centerline{$^{1}$ \it{L.D. Landau Institute for Theoretical Physics 
of Russian Academy of Sciences}}
\centerline{\it 142432 Chernogolovka, pr. Ak. Semenova 1A}
\centerline{$^{2}$ \it{V.A. Steklov Mathematical Institute 
of Russian Academy of Sciences}}
\centerline{\it 119991 Moscow, Gubkina str. 8}
}

\maketitle

\begin{abstract}
The paper is devoted to the questions connected with the investigation 
of the S.P. Novikov problem of the description of the geometry of level 
lines of quasiperiodic functions on a plane with different numbers of 
quasiperiods. We consider here the history of the question, the current 
state of research in this field, and a number of applications of this 
problem to various physical problems. The main attention is paid to the 
applications of the results obtained in the field under consideration 
to the theory of transport phenomena in electron systems.
\end{abstract}

\section{Introduction.}
\setcounter{equation}{0}

 In this paper we will mainly consider the formulation of problems,
research and results related to the S.P. Novikov problem, set
in the early 1980s (see \cite{MultValAnMorseTheory}) and having
relation in fact to many areas of research, in particular, such as 
the theory of dynamical systems, the theory of quasiperiodic functions,
the theory of topological phenomena in the physics of condensed matter, 
the theory of transport phenomena in systems of various dimensions, etc.
In the very first formulation, the problem of S.P. Novikov was connected 
with the description of the levels of multivalued functions on manifolds, 
but it has the same natural formulation also in the language of the theory 
of dynamical systems. Most of our paper will be connected with the  
investigations of the Novikov problem which relate to dynamical systems 
arising on complex Fermi surfaces in metals in the presence of an external 
magnetic field. As is well known, the description of the trajectories of 
such systems can be effectively reduced to the description of the level lines 
of height functions on periodic two-dimensional surfaces embedded in 
three-dimensional space. More precisely, the dynamics of electron states 
in a crystal in the presence of an external magnetic field is described 
by the system (see, e.g. \cite{Abrikosov,Kittel,Ziman})
\begin{equation}
\label{MFSyst}
{\dot {\bf p}} \,\,\, = \,\,\, {e \over c} \, 
\left[ {\bf v}_{gr} \times {\bf B} \right] \,\,\, = \,\,\,
{e \over c} \, \left[ \nabla \epsilon ({\bf p}) \times {\bf B} \right] 
\,\,\, , 
\end{equation}
where $\, {\bf p} \, $ is the quasimomentum of the electron state.
It is extremely important here that the space of quasimomenta represents
actually a three-dimensional torus $\, \mathbb{T}^{3} \, $, and not the 
Euclidean space $\, \mathbb{R}^{3} \, $, as in the case of free electrons.

 The space of electron states can also be represented as the space 
$\, \mathbb{R}^{3} \, $ provided that any two values of $\, {\bf p} \, $
that differ by a reciprocal lattice vector, define the same 
quantum-mechanical state. The function $\, \epsilon ({\bf p}) \, $ 
represents in this case a 3-periodic function in $\, \mathbb{R}^{3} \, $, 
and its level surfaces are two-dimensional 3-periodic surfaces in the same 
space. The reciprocal lattice $\, L^{*} \, $ in the quasimomentum space 
can be defined with the aid of its basis vectors
$\, {\bf a}_{1} $,  $\, {\bf a}_{2} $, $\, {\bf a}_{3} \, $, 
which are connected by the relations
$${\bf a}_{1} \,\,\, = \,\,\, 2 \pi \hbar \,\, 
{{\bf l}_{2} \times {\bf l}_{3} \over ({\bf l}_{1}, {\bf l}_{2}, {\bf l}_{3})} 
\,\,\, ,  \quad
{\bf a}_{2} \,\,\, = \,\,\, 2 \pi \hbar \,\, 
{{\bf l}_{3} \times {\bf l}_{1} \over ({\bf l}_{1}, {\bf l}_{2}, {\bf l}_{3})} 
\,\,\, ,  \quad
{\bf a}_{3} \,\,\, = \,\,\, 2 \pi \hbar \,\, 
{{\bf l}_{1} \times {\bf l}_{2} \over ({\bf l}_{1}, {\bf l}_{2}, {\bf l}_{3})} $$
with the basis vectors
$\, ( {\bf l}_{1}, \, {\bf l}_{2}, {\bf l}_{3} ) \, $ of the direct lattice 
of the crystal. As is not hard to see, the exact phase space then represents 
a torus
$$\mathbb{T}^{3} \,\,\, = \,\,\, \mathbb{R}^{3} / L^{*} \,\,\, , $$
obtained from the complete $\, {\bf p}$ - space with the aid of the 
factorization by the vectors
$$m_{1} \, {\bf a}_{1} \,\,\, + \,\,\, m_{2} \, {\bf a}_{2} \,\,\, + \,\,\,
m_{3} \, {\bf a}_{3} \,\,\, ,  \quad  \quad
m_{1}, \, m_{2}, \, m_{3} \,\, \in \,\, \mathbb{Z} $$

 The system (\ref{MFSyst}) is a Hamiltonian system with the Hamiltonian 
$\, H \, = \, \epsilon ({\bf p}) \, $ and the Poisson bracket
$$\{ p_{1} \, , \, p_{2} \} \,\, = \,\, {e \over c} \, B^{3} \,\,\, , \quad
\{ p_{2} \, , \, p_{3} \} \,\, = \,\, {e \over c} \, B^{1} \,\,\, , \quad
\{ p_{3} \, , \, p_{1} \} \,\, = \,\, {e \over c} \, B^{2} $$ 
 
 It is easy to see that the consequence of this fact is, in particular, 
the conservation of the quantity $\, \epsilon ({\bf p}) \, $, as well as 
the projection of the quasimomentum on the direction $\, {\bf B} \, $, along 
trajectories of the system. As a consequence, geometrically the trajectories 
of the system (\ref{MFSyst}) in $\, {\bf p}$ - space are actually given by the 
intersections of surfaces of constant energy
$$\epsilon ({\bf p}) \,\,\, = \,\,\, {\rm const} $$
with planes orthogonal to the magnetic field. It is also easy to see that due 
to this property the problem under consideration has two natural interpretations, 
namely, the problem of describing the trajectories of a dynamical system with 
an ambiguous conservation law on a compact surface and the problem of describing 
the level lines of a quasiperiodic function on a plane with three quasiperiods.
It must be said that both these interpretations play an extremely important role 
in the study of the problem.

 The important role of the geometry of the trajectories of the system 
(\ref{MFSyst}) in the behavior of the electrical conductivity of metals 
in strong magnetic fields (magnetoconductivity) was first discovered by the 
school of I.M. Lifshits (I.M. Lifshits, M.Ya. Azbel, M.I. Kaganov, V.G. Peschanskii) 
in the late 1950s - early 1960s (see
\cite{lifazkag,lifpes1,lifpes2,lifkag1,lifkag2,lifkag3,etm,KaganovPeschansky}). 
We note at once that in the theory of normal metals usually only the single 
energy level $\, \epsilon_{F} \, $ (the Fermi level) is important for most 
processes occurring in metal, while all levels far from the Fermi level are 
always either completely filled, or empty, and do not affect the ongoing processes.
As a consequence, in the theory of galvanomagnetic phenomena, only the trajectories 
of the system (\ref{MFSyst}), which lie on the Fermi surface
$$S_{F}:  \quad \epsilon ({\bf p}) \,\,\, = \,\,\, \epsilon_{F} \quad , $$
are interesting, and it is the complexity of the Fermi surface that determines 
the features of electron transport phenomena in strong magnetic fields.

 Thus, in the paper \cite{lifazkag} there was indicated the principal difference 
in the behavior of the magnetoconductivity in strong magnetic fields in 
the presence of only closed trajectories of the system (\ref{MFSyst}) on the 
Fermi surface in $\, {\bf p}$ - space and in the presence of periodic open 
trajectories in $\, {\bf p}$ - space on this surface. In both cases, the 
asymptotic behavior of the conductivity tensor can be represented as a regular 
series in inverse powers of $\, B \, $, and, in particular, has the following 
form in the presence of only closed trajectories on the Fermi surface in 
$\, {\bf p}$ - space:
\begin{equation}
\label{Closed}
\sigma^{ik} \,\,\,\, \simeq \,\,\,\,
{n e^{2} \tau \over m^{*}} \, \left(
\begin{array}{ccc}
( \omega_{B} \tau )^{-2}  &  ( \omega_{B} \tau )^{-1}  &
( \omega_{B} \tau )^{-1}  \cr
( \omega_{B} \tau )^{-1}  &  ( \omega_{B} \tau )^{-2}  &
( \omega_{B} \tau )^{-1}  \cr
( \omega_{B} \tau )^{-1}  &  ( \omega_{B} \tau )^{-1}  &  *
\end{array}  \right)  ,   \quad \quad
\omega_{B} \tau \,\, \rightarrow \,\, \infty 
\end{equation}

 In the formula (\ref{Closed}) the value $\, n \, $ represents 
the concentration of charge carriers in the metal, and the value 
$\, m^{*} \, $ determines the order of the effective mass of the electron 
in the crystal. The time $\, \tau \, $ plays the role of the mean free time 
of an electron and depends on the purity as well as the temperature of 
the crystal. The value $\, \omega_{B} \, = \, e B / m^{*} c \, $ 
has the meaning of the cyclotron frequency of the electron in the crystal, 
it should be noted here that, unlike the case of the free electron gas, 
the cyclotron frequency is defined here only for closed trajectories of 
the system (\ref{MFSyst}) and coincides with the parameter 
$\, \omega_{B} \, $ only in order of magnitude. The sign $\, \simeq \, $ 
expresses here the equality up to a dimensionless coefficient (of order 1) 
and the notation $\, * \, $ means also some constant. Here and in what follows 
we will always assume that the axis $\, z \, $ in our considerations is chosen 
along the direction of the magnetic field. It is easy to see that the formula 
(\ref{Closed}) coincides essentially with the analogous formula for the free 
electron gas, differing from it only by possible numerical parameters.

 A completely different situation arises in the presence of periodic open 
trajectories on the Fermi surface in $\, {\bf p}$ - space. As was shown in 
\cite{lifazkag}, after the choice of the axis $\, x \, $ along the average 
direction of the periodic trajectory in $\, {\bf p}$ - space, the principal 
term of the asymptotic expansion of the tensor $\, \sigma^{ik} (B) \, $ can be 
represented in the form
\begin{equation}
\label{Periodic}
\sigma^{ik} \,\,\,\, \simeq \,\,\,\,
{n e^{2} \tau \over m^{*}} \, \left(
\begin{array}{ccc}
( \omega_{B} \tau )^{-2}  &  ( \omega_{B} \tau )^{-1}  &
( \omega_{B} \tau )^{-1}  \cr
( \omega_{B} \tau )^{-1}  &  *  &  *  \cr
( \omega_{B} \tau )^{-1}  &  *  &  *
\end{array}  \right)  ,   \quad \quad
\omega_{B} \tau \,\, \rightarrow \,\, \infty 
\end{equation}

 It can be seen that the electric conductivity tensor here has a strong 
anisotropy in the plane orthogonal to $\, {\bf B} $ in strong magnetic fields, 
which makes its behavior fundamentally different from the case of the free 
electron gas. It is easy to see that the formula (\ref{Periodic}) allows also 
to measure the average direction of periodic trajectories in $\, {\bf p}$ - space 
as the direction of the greatest suppression of conductivity in the plane 
orthogonal to $\, {\bf B} $. Let us note here that both the closed 
and periodic trajectories in $\, {\bf p}$ - space are represented by
closed trajectories of the system (\ref{MFSyst}) in the torus
$\, \mathbb{T}^{3} \, $. It can thus be seen that for the description of 
galvanomagnetic phenomena in metals, not only the shape of the trajectories 
of (\ref{MFSyst}) on the Fermi surface is important, but also their homology 
classes under the embedding in the torus $\, \mathbb{T}^{3} $. 

 Quasiclassical trajectories of the system (\ref{MFSyst}) correspond also 
to quasiclassical trajectories (electron packets) in the $\, {\bf x}$ - space, 
which are determined from the system
$${\dot {\bf x}} \,\,\, = \,\,\, {\bf v}_{gr} ({\bf p}) 
\,\,\, = \,\,\, \nabla \epsilon ({\bf p}) $$

 The electron trajectories in $\, {\bf x}$ - space do not coincide in common 
with the trajectories of the system (\ref{MFSyst}), in particular, they are 
not plane. However, their shape correlates quite strongly with the shape 
of the trajectories of (\ref{MFSyst}). For example, the projections of the 
electron trajectories in the $\, {\bf x}$ - space onto the plane, orthogonal 
to $\, {\bf B} \, $, are similar to the corresponding trajectories of the system 
(\ref{MFSyst}), rotated by $90^{\circ}$. As already noted above, the shape 
of the trajectories of the system (\ref{MFSyst}) becomes important in the limit 
$\, \omega_{B} \tau \gg 1 \, $, which can be formulated as the condition 
that the electron turns many times along a closed trajectory or passes 
a distance much greater than the size of the Brillouin zone in 
$\, {\bf p}$ - space between two scattering acts. 
 
 In the papers \cite{lifpes1,lifpes2} examples of open trajectories 
of a more general form on Fermi surfaces of different shapes were considered.
The trajectories considered in \cite{lifpes1,lifpes2} are not periodic in 
general case, however, they also have an average direction in the plane 
orthogonal to $\, {\bf B} $, which also leads to a sharp anisotropy of the 
electric conductivity tensor in the same plane. It must be said, however, 
that the analytic properties of the electric conductivity tensor are generally 
more complex here in comparison with the case of periodic trajectories 
(see e.g. \cite{KaganovPeschansky,AnProp}). 

 The problem of the complete classification of various types of trajectories 
of system (\ref{MFSyst}) was first set by S.P. Novikov in the work 
\cite{MultValAnMorseTheory} and was actively investigated in his topological 
school during the last decades. At present, the Novikov problem of the 
classification of trajectories of (\ref{MFSyst}) 
(with an arbitrary dispersion law) has been studied in many details, 
and in particular, rather profound results have been obtained, which have 
already found applications in the theory of solids. 
 
 Let us note at once that the most important part of the results obtained 
in the investigation of Novikov's problem was the description of stable open 
trajectories of the system (\ref{MFSyst}) for an arbitrary dispersion law 
(A.V. Zorich, I.A. Dynnikov). The remarkable geometric properties of such 
trajectories made it possible to define important topological characteristics 
(topological quantum numbers) observable in studies of conductivity of normal 
metals, which, in particular, provide a convenient tool for determining the 
orientation of the crystal lattice in such studies 
(S.P. Novikov, A.Ya. Maltsev). We also note here that the description 
of the geometric properties of stable open trajectories of system 
(\ref{MFSyst}) allows, in addition, to approach more strictly the description 
of the analytic properties of conductivity in strong magnetic fields in the 
presence of trajectories of this type.

 At the same time, a detailed study of the system (\ref{MFSyst}) has also led 
to the discovery of new, rather nontrivial, types of trajectories of such 
systems whose properties are the subject of active study at the present time.
Let us note here that these trajectories exhibit very interesting (chaotic) 
properties both from the geometric point of view and from the point of view 
of the description of transport phenomena in normal metals when they appear.

 We must now say that the applications of the problem of describing the 
trajectories of the system (\ref{MFSyst}) are not really limited to transport 
phenomena in normal metals in strong magnetic fields. For example, the problem 
of describing the trajectories of systems similar to (\ref{MFSyst}) also arises 
in the description of transport phenomena in two-dimensional electron systems 
placed in artificially created quasiperiodic (super)potentials in the presence 
of an external magnetic field. The Novikov problem is formulated here as 
the problem of describing the geometry of the level lines of a quasiperiodic 
function on a plane with a fixed number of quasiperiods. It is not difficult 
to see here that the problem of describing the level curves of a function with 
three quasiperiods (the first case following the periodic one) coincides 
in reality with the Novikov problem in its formulation given above.
Thus, all the results obtained in the study of the trajectories of the system 
(\ref{MFSyst}) are also transferred to transport phenomena in two-dimensional 
electron systems in superpotentials with three quasiperiods. We must note here 
that, unlike the situation with normal metals, the parameters of such systems 
are controlled in this case, which makes it possible to implement any of the 
cases of behavior of the system (\ref{MFSyst}) which is interesting to us.  
 
 The problem of describing the geometry of the level lines of functions 
with a larger number of quasiperiods is actually much more complicated.
For example, for the case of functions with four quasiperiods, a rather 
serious result has now been obtained (S.P. Novikov, I.A. Dynnikov), which 
distinguishes an important class of potentials that have topologically regular 
open level lines analogous to stable open trajectories of the system 
(\ref{MFSyst}) in the three-dimensional case. But on the whole, the Novikov 
problem for four quasiperiods has been studied much less compared to the case 
of three quasiperiods.

 Here we try to give the most complete overview of the results obtained 
to date and describe, as much as possible, the range of possible applications 
of the Novikov problem in physical problems.

\section{Stable open trajectories and angular conductivity diagrams 
for normal metals.}
\setcounter{equation}{0}

 In this chapter we will consider stable open trajectories of the system 
(\ref{MFSyst}) and describe the main properties of transport phenomena in 
metals in strong magnetic fields in the presence of such trajectories on 
the Fermi surface. We say at once that under the stability of the open 
trajectories of the system (\ref{MFSyst}) we mean here the preservation 
of such trajectories, as well as their geometric properties, for all small 
variations in the direction of $\, {\bf B} \, $ or the energy level
$\, \epsilon ({\bf p}) \, = \, \epsilon_{0} \, $.
We also note that we consider the system (\ref{MFSyst}) in the covering 
space of quasimomenta, where geometrically its trajectories are given by 
the intersections of surfaces of constant energy and planes orthogonal to 
the magnetic field. As we said above, the dispersion relation 
$\, \epsilon ({\bf p}) \, $ is assumed here to be an arbitrary smooth 
3-periodic function in the $\, {\bf p}$ - space, with periods equal to the 
vectors of the reciprocal lattice.

 Let us describe here two important properties of stable open trajectories 
of the system (\ref{MFSyst}), which follow from the results obtained in the  
papers \cite{zorich1,dynn1992,dynn1}.

\vspace{2mm}

\begin{figure}[t]
\begin{center}
\includegraphics[width=\linewidth]{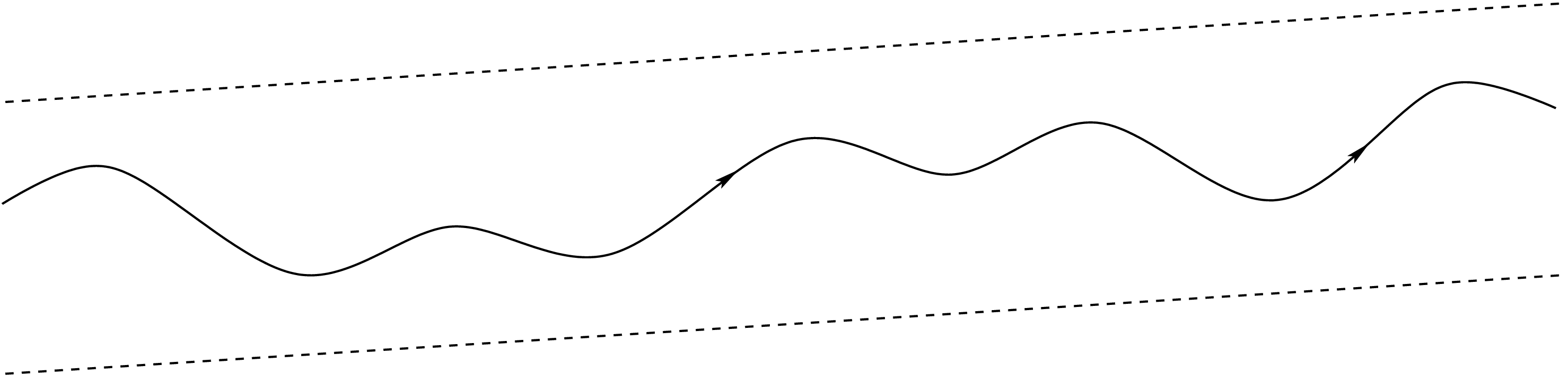}
\end{center}
\caption{The form of a stable open trajectory in the plane, orthogonal to 
$\, {\bf B} \, $, in the space of quasimomenta.}
\label{StableTr}
\end{figure}

1) All stable open trajectories of the system (\ref{MFSyst}) for a fixed 
direction of $\, {\bf B} \, $ lie in straight strips of finite width in planes 
orthogonal to $\, {\bf B} \, $, passing through them (Fig. \ref{StableTr}).

\vspace{1mm}

2) All stable open trajectories of the system (\ref{MFSyst}) for a fixed 
direction of $\, {\bf B} \, $ have the same mean direction in $\, {\bf p}$ - space 
defined by the intersection of the plane orthogonal to $\, {\bf B} \, $ with some 
integral (generated by two vectors of the reciprocal lattice) plane 
$\, \Gamma \, $, unchanged for all close directions of $\, {\bf B} \, $.

\vspace{2mm}

 In a more precise formulation, it follows from the papers 
\cite{zorich1,dynn1992} that the properties (1)-(2) hold for open trajectories 
of (\ref{MFSyst}) for the directions of $\, {\bf B} \, $, sufficiently close 
to rational directions, while it follows from the paper \cite{dynn1} that 
properties (1)-(2) hold for open trajectories that are stable with respect 
to variations of the energy level
$\, \epsilon ({\bf p}) \, = \, \epsilon_{0} \, $.
It is not difficult to see here that to satisfy conditions (1)-(2) it is 
sufficient to require either the stability of trajectories with respect to 
small rotations of the direction of $\, {\bf B} \, $, or stability with 
respect to variations of the energy level. Let us also note here that property 
(1) was first expressed by S.P. Novikov in the form of a conjecture, and was 
thus proved later for stable open trajectories of (\ref{MFSyst}).

 Conditions (1) and (2) play an important role for transport phenomena in 
metals and served as the basis for introducing in \cite{PismaZhETF} 
(see also \cite{UFN}) important topological characteristics 
(topological quantum numbers) observable in magnetic conductivity 
in the presence of stable open trajectories on the Fermi surface.
Thus, the fulfillment of conditions (1) and (2) leads to a strong anisotropy of 
the conductivity in the plane orthogonal to $\, {\bf B} \, $ in the limit 
$\, \omega_{B} \tau \gg 1 \, $, which makes it possible to measure the average 
direction of stable open trajectories with direct measurement of conductivity 
in strong magnetic fields. It should be noted here that the analytic 
behavior of the conductivity is, in general, more complicated than the 
dependence (\ref{Periodic}) in this situation (see \cite{AnProp}), but in any 
case, we have here the relations
\begin{equation}
\label{GeneralLimit}
\sigma^{ik} \,\,\,\, \simeq \,\,\,\,
{n e^{2} \tau \over m^{*}} \, \left(
\begin{array}{ccc}
o (1)  &  o (1)  & o (1)  \cr
o (1)  &  *  &  *  \cr
o (1)  &  *  &  *
\end{array}  \right)  ,   \quad \quad
\omega_{B} \tau \,\, \rightarrow \,\, \infty 
\end{equation}
provided that the axis $\, x \, $ coincides with the mean direction of the 
open trajectories in the $\, {\bf p}$ - space. Thus, the mean direction of 
stable open trajectories in $\, {\bf p}$ - space is directly observable as 
the direction of the largest suppression of conductivity in the plane orthogonal 
to $\, {\bf B}$ in strong magnetic fields. Variations of the direction of the 
magnetic field within the stability zone for the given family of open 
trajectories determine, in this case, the direction of the integral plane 
$\, \Gamma \, $ in $\, {\bf p}$ - space associated with this family according 
to condition (2).

  We note here that the plane $\, \Gamma \, $ is integral in the space 
of quasimomenta, which means that it is generated by some two vectors of 
the reciprocal lattice. In particular, it does not have to coincide in the 
general case with any of the crystallographic planes in the coordinate space.
Instead, it can be given in coordinate space by the relation
$$M_{1} \, \left( {\bf x} , {\bf l}_{1} \right) \,\, + \,\,
M_{2}   \, \left( {\bf x} , {\bf l}_{2} \right) \,\, + \,\,
M_{3}   \, \left( {\bf x} , {\bf l}_{3} \right) 
\,\,\, = \,\,\, 0 \,\,\, , \quad \quad 
M_{1}, M_{2}, M_{3} \, \in \, \mathbb{Z} $$
and, thus, is orthogonal to one of the integer crystallographic directions.
The irreducible integer triple $\, (M_{1}, M_{2}, M_{3}) \, $ represents
here a topological characteristic of the corresponding family of stable 
open trajectories that is directly observable in conductivity measurements 
in strong magnetic fields. Integral triples 
$\, (M^{\alpha}_{1}, M^{\alpha}_{2}, M^{\alpha}_{3}) \, $
for the complete set of all different Stability Zones $\, \Omega_{\alpha} \, $
in the space of directions of $\, {\bf B} \, $ were called in \cite{PismaZhETF}
the topological quantum characteristics (topological quantum numbers) 
observable in the conductivity of normal metals.

 It can thus be seen that the measurement of the conductivity for different 
directions of $\, {\bf B} \, $ within the same Stability Zone 
$\, \Omega_{\alpha} \, $ allows us to determine quite accurately some 
(known) crystallographic direction in the coordinate space.
The same conductivity measurement for the directions of $\, {\bf B} $ lying 
in two different Stability Zones (with different topological quantum numbers) 
makes it possible to completely determine the orientation of the crystal 
lattice of a single crystal. Let us note here that such a method of determining
the orientation of a single crystal sample is significantly more convenient than, 
for example, the measurement of the exact angular diagram of conductivity, 
since (as we shall see below) the exact boundaries of the Stability Zones 
are actually difficult to observe in direct conductivity measurements 
(see, e.g. \cite{AnProp,CyclRes}), and also because precise determination 
of the theoretical boundaries of the Stability Zones for a given dispersion 
relation is also a serious task in the general case (see \cite{DeLeoPhysB}).

 Stability Zones represent domains with piecewise smooth boundaries 
in the space of directions of $\, {\bf B} \, $ (on the unit sphere 
$\, \mathbb{S}^{2}$). The full Stability Zone $\, \Omega_{\alpha} \, $
can be defined as a complete domain on the sphere $\, \mathbb{S}^{2} \, $
such that for any direction $\, {\bf B} \, \in \, \Omega_{\alpha} \, $
there are stable open trajectories on the Fermi surface that correspond 
to the same topological quantum numbers
$\, (M^{\alpha}_{1}, M^{\alpha}_{2}, M^{\alpha}_{3}) $.
It is easy to see that the full Stability Zone is invariant under 
the substitution $\, {\bf B} \, \rightarrow \, - {\bf B} \, $ and often 
consists of two opposite connected components on the unit sphere.

 The addition to the union of the bands $\, \Omega_{\alpha} \, $
on the unit sphere
$$\hat{S}^{2} \,\,\, = \,\,\, \mathbb{S}^{2} \setminus
\cup \Omega_{\alpha} $$
is by definition the set of directions of $\, {\bf B} \, $ for which 
stable open trajectories are not present on the Fermi surface.
At the same time, however, the set $\, \hat{S}^{2} \, $
can contain directions of $\, {\bf B} $, corresponding to the appearance 
of unstable open trajectories of the system (\ref{MFSyst}) on the Fermi 
surface. Among such directions, we should especially note the directions 
of $\, {\bf B} $, which lead to the appearance of periodic open trajectories 
on the Fermi surface that are unstable for 
$\, {\bf B} \, \in \, \hat{S}^{2} \, $.
The presence of a large number of such directions of
$\, {\bf B} \, $ on the set $\, \hat{S}^{2} \, $ is in fact a consequence 
of the topological structure of the system (\ref{MFSyst}) for 
$\, {\bf B} \, \in \, \Omega_{\alpha} \, $, which we briefly describe below.

 The topological structure of the system (\ref{MFSyst}) in the presence 
of stable open trajectories was described in the papers \cite{zorich1,dynn1} 
and is based on a property that can be called ``topological integrability'' 
for systems of this type. Namely, consider the system (\ref{MFSyst}) on a
fixed surface $\, \epsilon ({\bf p}) \, = \, {\rm const} \, $. Let us assume 
that the direction of $\, {\bf B} \, $ has maximal irrationality and remove 
from this surface all closed trajectories of the system (\ref{MFSyst}) 
(homologous to zero in the torus $\, \mathbb{T}^{3} $). It is not difficult 
to see that the remainder of the surface will be the carrier of the open 
trajectories of the system (\ref{MFSyst}). As follows from the works 
\cite{zorich1,dynn1}, in the presence of stable open trajectories of 
the system (\ref{MFSyst}) on the surface, each connected component
of such a carrier represents a two-dimensional torus $\, \mathbb{T}^{2} \, $
with contractible holes embedded in the torus $\, \mathbb{T}^{3} $. 
  
\begin{figure}[t]
\begin{center}
\includegraphics[width=0.95\linewidth]{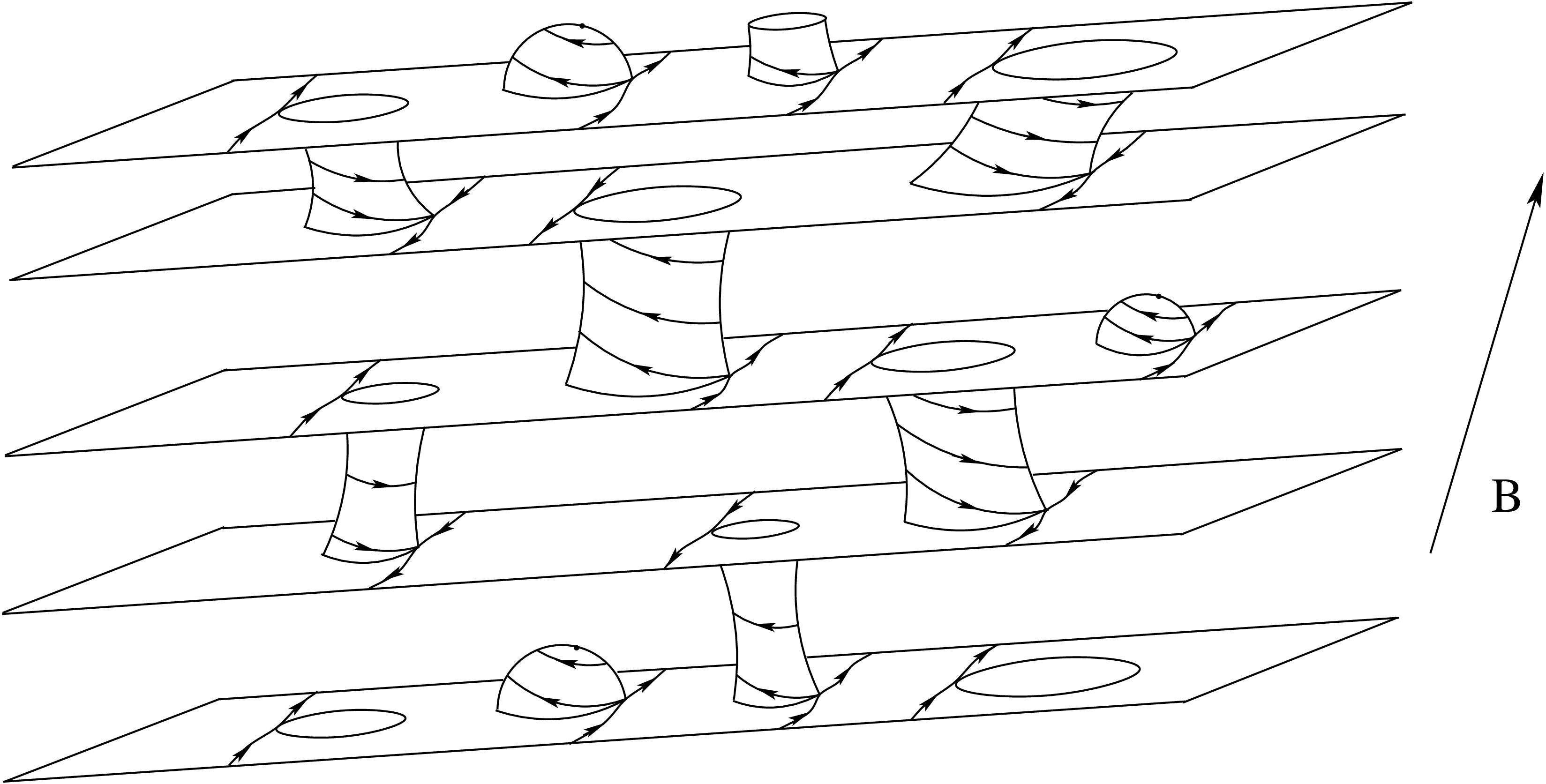}
\end{center}
\caption{Structure of a connected component of the Fermi surface carrying 
stable open trajectories of the system (\ref{MFSyst}) in the covering 
$\, {\bf p}$ - space.}
\label{StableTrFermiSurf}
\end{figure}

  Returning to the covering $\, {\bf p}$ - space, we can thus say that any 
connected component of the energy (Fermi) surface carrying stable open 
trajectories of the system (\ref{MFSyst}) has a well-defined topological 
structure, defined by the trajectories of the system (\ref{MFSyst}).
Namely, each such connected component represents a union of integral 
(periodically deformed) planes in the $\, {\bf p}$ - space connected by 
parts formed by cylinders of closed trajectories of the system (\ref{MFSyst}).
It can be noted here also that for almost any real dispersion law the 
corresponding parts will in fact be simple cylinders of closed trajectories 
bounded by singular trajectories on their bases (Fig. \ref{StableTrFermiSurf}).

 The results of investigations of the Novikov problem in solid state 
theory are most significant in cases when the dispersion relation for the 
electron in a crystal has a rather complex form. In particular, this refers 
to the shape of the Fermi surface, which is also assumed here to be quite 
complicated. We can say, for example, that a metal has a complex Fermi surface 
$\, S_{F} $, if it has rank 3, where the rank of the surface is determined by 
the dimension of the image of the mapping
$$H_{1} (S_{F}) \,\,\, \rightarrow \,\,\, H_{1} (\mathbb{T}^{3}) $$
 
 It can also be seen that a connected Fermi surface of rank 3 must 
necessarily have genus $\, g \, \geq \, 3 $.

 Thus, according to the introduced terminology, a complex connected 
component of the Fermi surface, having the structure shown at 
Fig. \ref{StableTrFermiSurf}, must contain at least two nonequivalent 
families of parallel planes connected by cylinders of closed trajectories 
of both ``electron'' and ``hole'' type. In the general case, as is not hard 
to see, the number of nonequivalent integral planes in this representation 
of the Fermi surface is necessarily even, and the corresponding planes can be 
divided into two classes in accordance with the direction of motion 
(``forward'' or ``back'') along the corresponding open trajectories in
$\, {\bf p}$ - space. It can also be noted that for the overwhelming number 
of real crystals, the number of nonequivalent integral planes in the described 
representation is exactly two, since a larger number of such planes corresponds 
to rather large genera of the Fermi surface. Based on the form of real dispersion 
relations, it is this situation, therefore, that should be considered typical 
when stable open trajectories of the system (\ref{MFSyst}) appear on the Fermi 
surface. We also note here that the structure shown at 
Fig. \ref{StableTrFermiSurf} has a purely topological character and can be 
visually much more complicated for real Fermi surfaces.
 
 It is not difficult to see that the presence of at least one pair of carriers 
of stable open trajectories excludes the appearance of open trajectories of the 
system (\ref{MFSyst}) having a different form, and that all the present open 
trajectories have the same direction in the $\, {\bf p}$ - space in this case.
We note that this property is also true for Fermi surfaces consisting of several 
connected components (in the torus $\, \mathbb{T}^{3}$), provided that these 
components do not intersect each other. This circumstance is especially important 
for Fermi surfaces of real crystals, which usually consist of several components, 
determined by different dispersion relations (the property of non-intersection of 
different components of the Fermi surface is, as a rule, preserved in this case).
This property of stable open trajectories for the full Fermi surface was called 
the Topological Resonance in \cite{BullBrazMathSoc,JournStatPhys} and plays an 
important role in describing angular diagrams for conductivity in real conductors.
In particular, this property excludes the possibility of crossing two Stability 
Zones with different topological quantum numbers on the angular diagram.

\begin{figure}[t]
\begin{center}
\includegraphics[width=\linewidth]{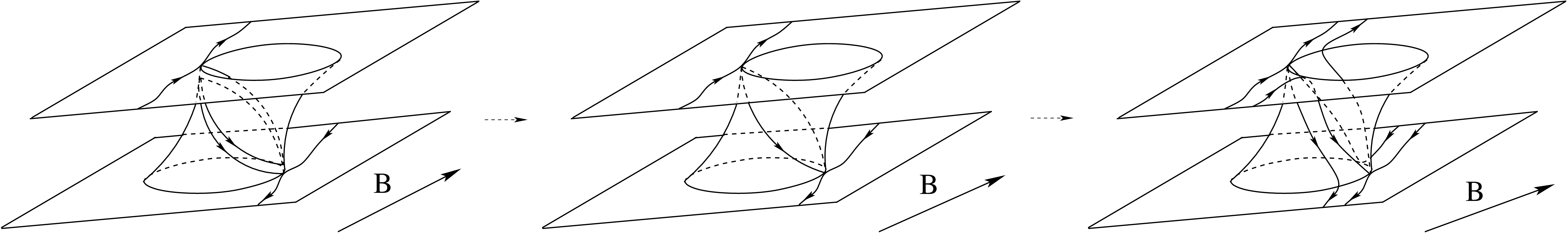}
\end{center}
\caption{The vanishing of the height of one of the cylinders of closed 
trajectories of the system (\ref{MFSyst}), followed by a ``jump'' of trajectories 
from one integral plane to another when crossing the boundary of a Stability Zone 
on the angular diagram.}
\label{Cylind}
\end{figure}

 It is easy to see also, that in the situation described above the open 
trajectories of the system (\ref{MFSyst}) exist on all the presented integral 
planes for all directions of $\, {\bf B} $, except possibly for the direction 
orthogonal to the plane $\, \Gamma_{\alpha} $ (if this direction belongs to the 
Zone $\, \Omega_{\alpha} $). It can also be seen that the described structure 
of the Fermi surface is preserved under rotations of the direction of
$\, {\bf B} \, $ as long as all the cylinders of closed trajectories connecting 
the integral planes remain unbroken. The boundary of the corresponding Stability 
Zone on the angular diagram is thus determined by the condition that the height 
of one of the cylinders of closed trajectories described above (Fig. \ref{Cylind}) 
goes to zero.

\begin{figure}[t]
\begin{tabular}{lc}
\includegraphics[width=0.6\linewidth]{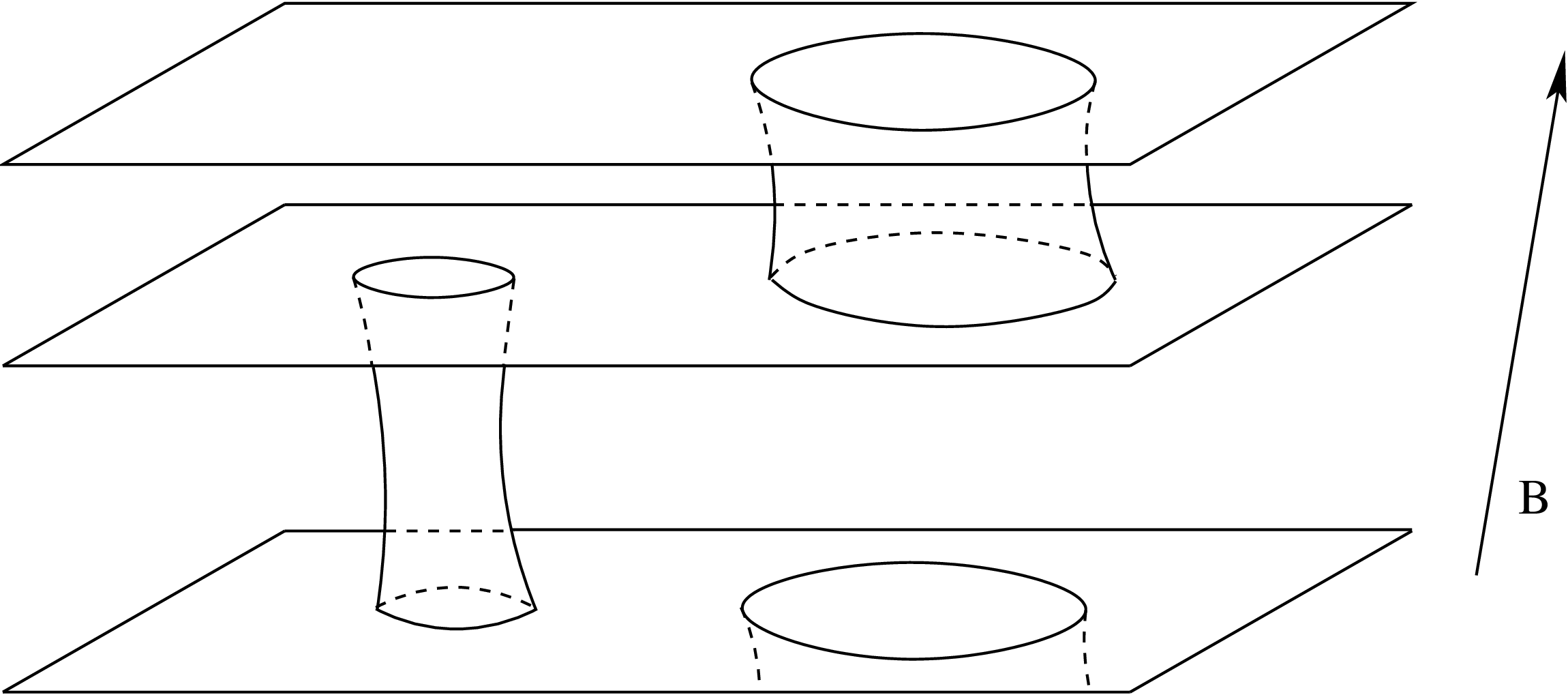}  &
\hspace{5mm}
\includegraphics[width=0.3\linewidth]{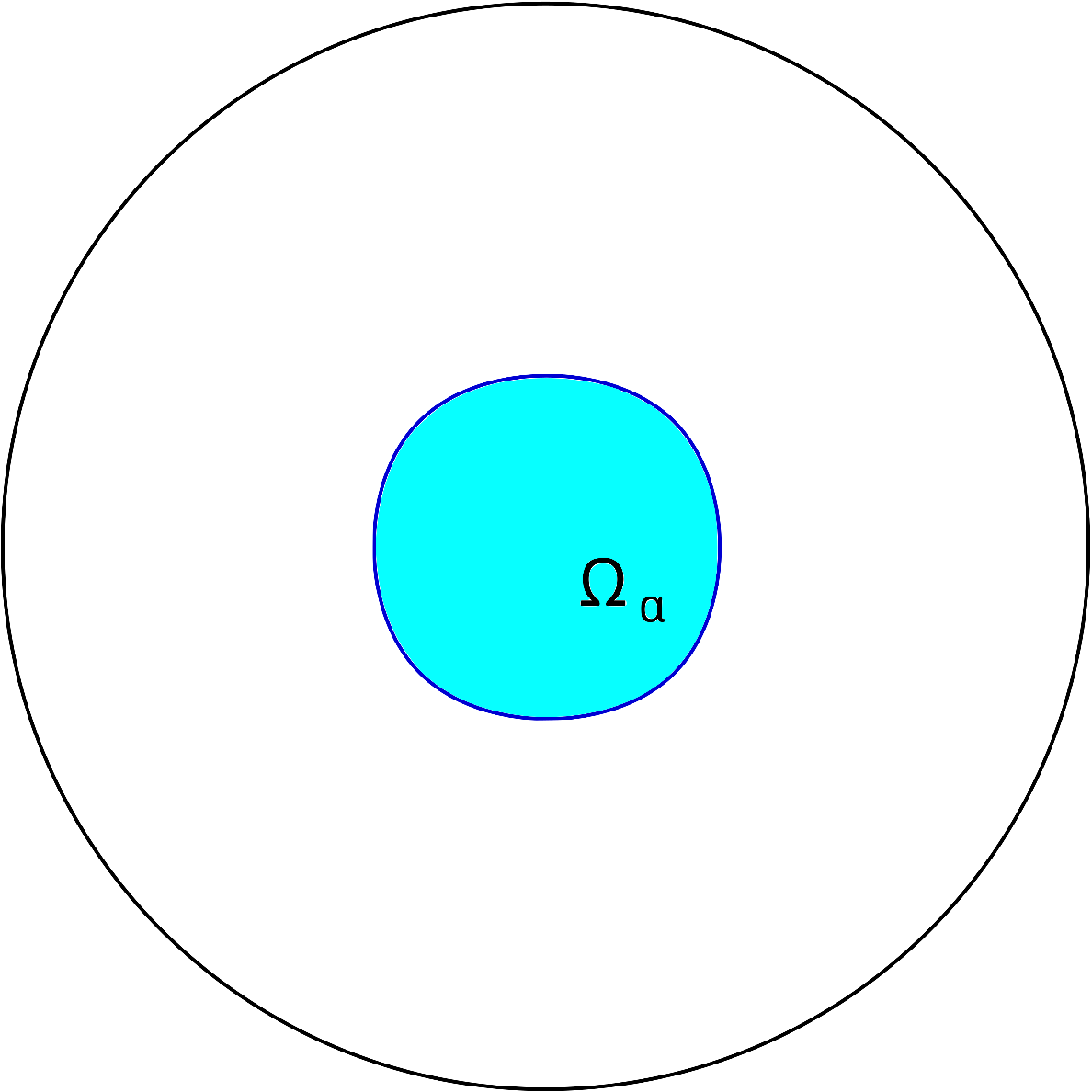}
\end{tabular}
\caption{The Fermi surface having a Stability Zone 
$\, \Omega_{\alpha} \, $ with a simple boundary.}
\label{SimpleFirstBound}
\end{figure}

\begin{figure}[t]
\begin{tabular}{lc}
\includegraphics[width=0.6\linewidth]{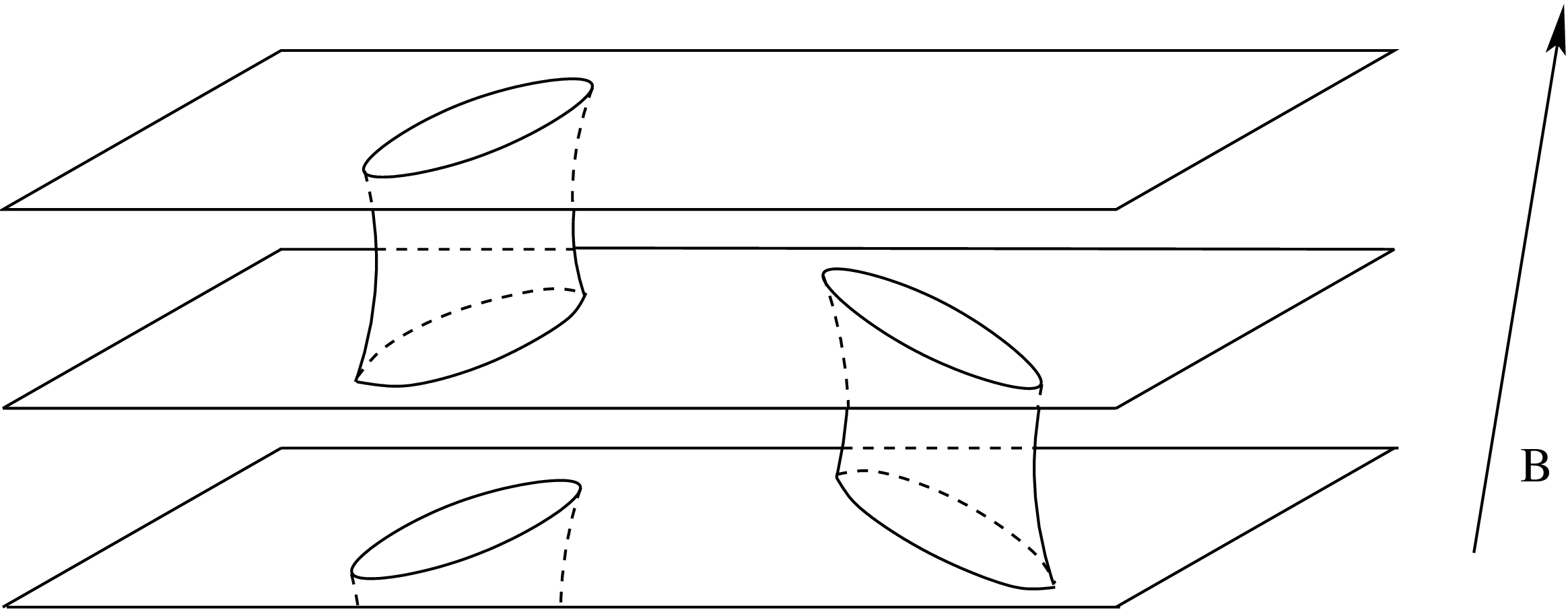}  &
\hspace{5mm}
\includegraphics[width=0.3\linewidth]{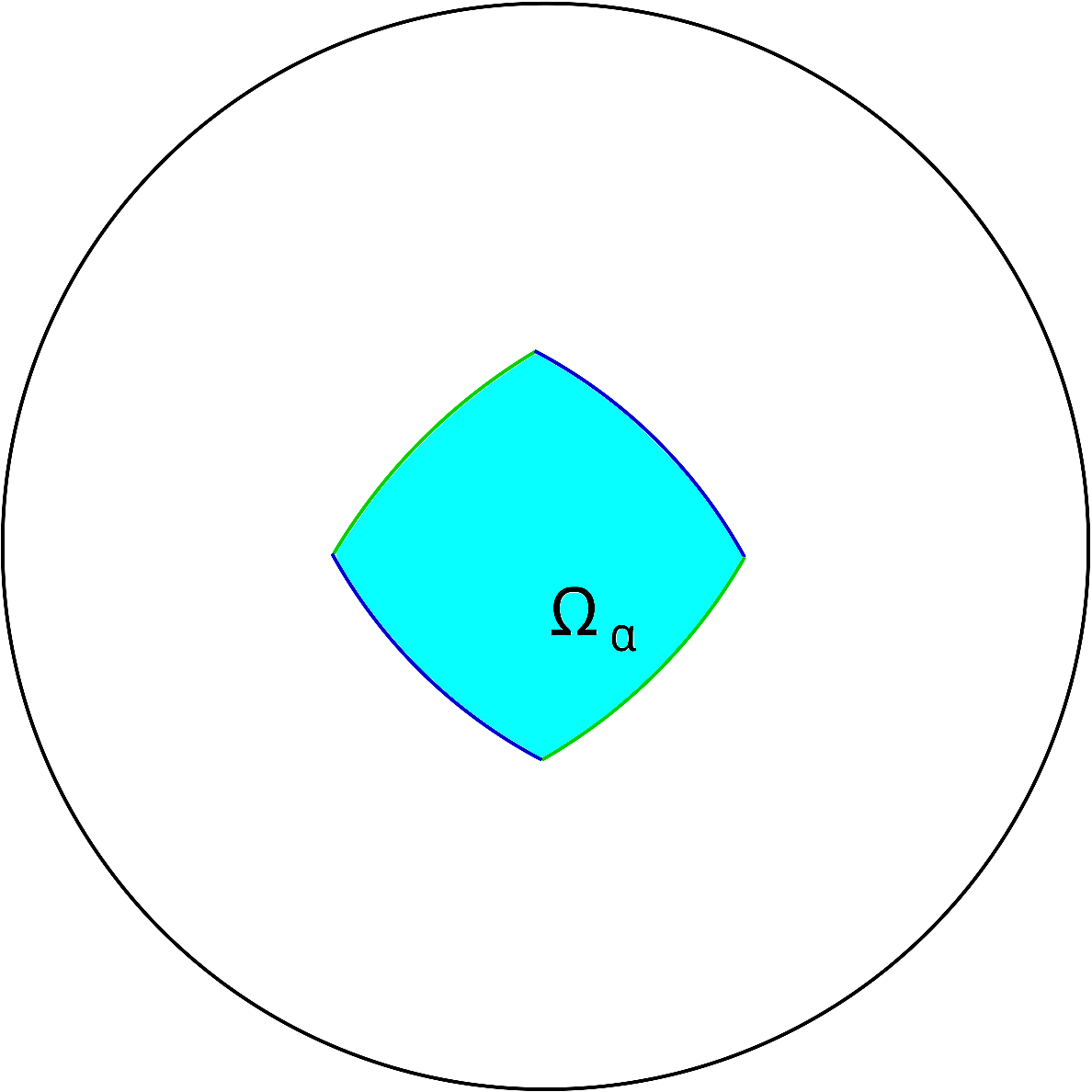}
\end{tabular}
\caption{The Fermi surface having a Stability Zone 
$\, \Omega_{\alpha} \, $ with a compound boundary.}
\label{CompoundFirstBound}
\end{figure}

\begin{figure}[t]
\begin{center}
\includegraphics[width=0.95\linewidth]{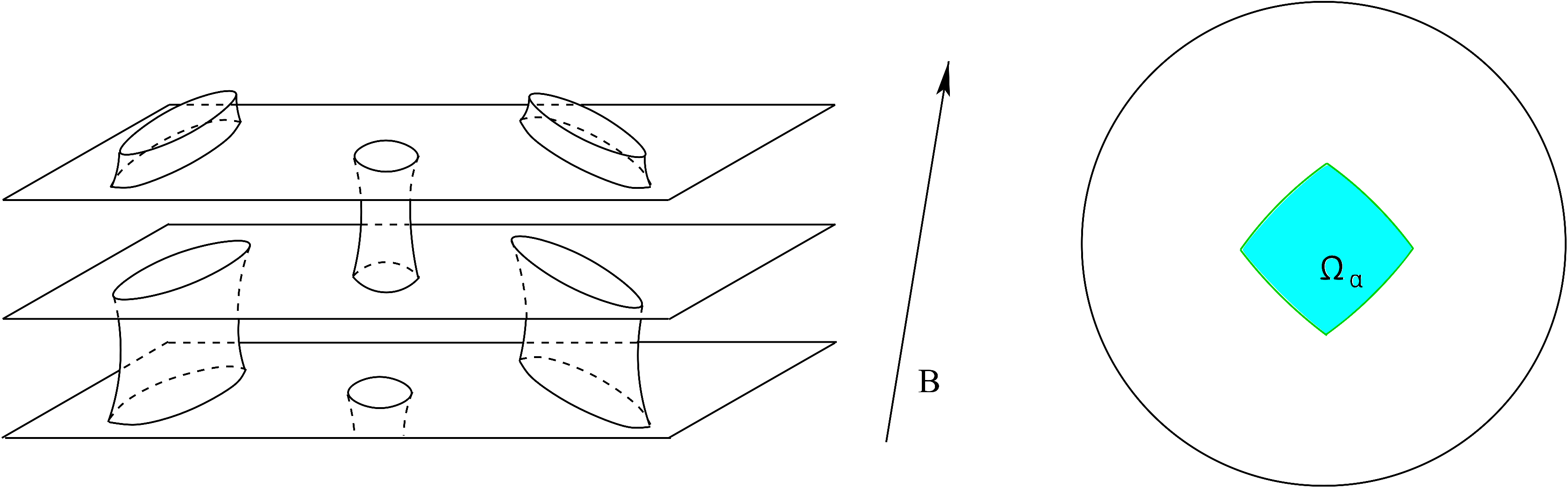}  
\end{center}
\caption{The Fermi surface having a Stability Zone 
$\, \Omega_{\alpha} \, $ with a compound boundary, determined by 
the disappearance of different cylinders of the same type in its 
different parts.}
\label{CompoundOneType}
\end{figure}

 It is not difficult to see that the total boundary of a Stability Zone on 
the angular diagram can be either ``simple'' and correspond to the disappearance 
of only one cylinder of closed trajectories (Fig. \ref{SimpleFirstBound}), or 
``compound'' and be determined by the disappearance of different cylinders of 
closed trajectories in its different parts (Fig. \ref{CompoundFirstBound}).
Besides that, it is also easy to see that in the case of a compound boundary 
of a Stability Zone we can have cases when the cylinders of different 
(electron and hole) types disappear in different parts of the boundary 
(Fig. \ref{CompoundFirstBound}), and when different cylinders of the same type 
disappear in its different parts (Fig. \ref{CompoundOneType}).

\begin{figure}[t]
\begin{center}
\includegraphics[width=0.95\linewidth]{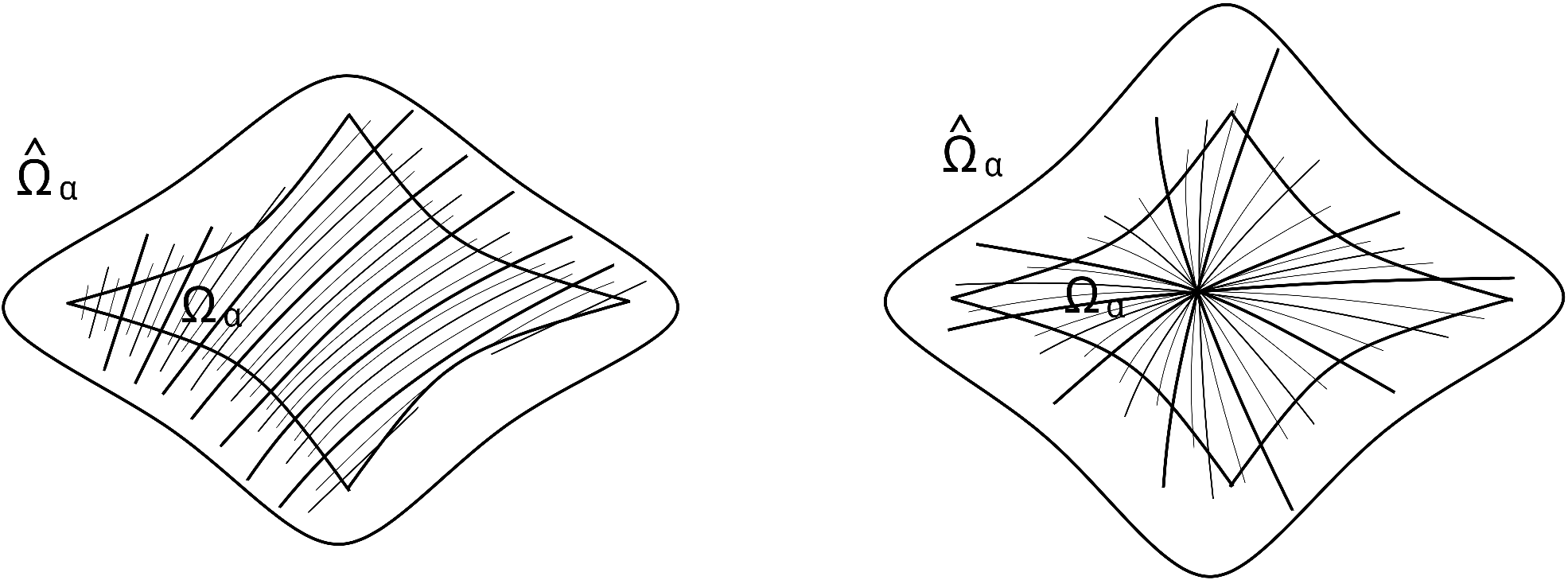}  
\end{center}
\caption{A schematic view of the experimental Stability Zones 
$\, \hat{\Omega}_{\alpha} \, $ observable in direct measurements of 
conductivity in strong magnetic fields (the ``nets'' of special directions 
of $\, {\bf B} $, corresponding to the appearance of periodic trajectories 
on the Fermi surface, are also shown).}
\label{ExpZone}
\end{figure}

 What also can be noted in the situation described above is that the 
representation of the Fermi surface, shown at Fig. \ref{StableTrFermiSurf}, 
allows actually to give an effective description of the trajectories of the 
system (\ref{MFSyst}) also immediately after crossing the boundary of the 
Stability Zone $\, \Omega_{\alpha} \, $ on the angular diagram. Indeed, 
after the intersection of the boundary of $\, \Omega_{\alpha} \, $ the 
trajectories can jump between the previous carriers of open trajectories, 
however, the Fermi surface is still divided into pairs of such carriers 
separated from each other by the remaining cylinders of closed trajectories. 
It can be seen in this case that the corresponding component is completely 
divided into closed trajectories if the intersection of the plane orthogonal 
to $\, {\bf B} $ with the plane $\, \Gamma_{\alpha} \, $ has an irrational 
direction, and can contain open periodic trajectories if this intersection 
has a rational direction in the $\, {\bf p}$ - space. It is also easy to see 
that the resulting closed trajectories are strongly elongated in one direction 
and have very long length, so that in the immediate vicinity of the boundary 
of $\, \Omega_{\alpha} \, $ we have the relation $\, T \, \geq \, \tau \, $,
where $\, T \, $ is the typical time of a turn along such trajectories.
As a consequence, trajectories of this type are indistinguishable from open 
trajectories from the experimental point of view, and the exact boundary of 
the Zone $\, \Omega_{\alpha} \, $ is actually unobservable in direct conductivity
measurements even in rather strong magnetic fields. Besides that, when approaching 
the boundary of the Zone $\, \Omega_{\alpha} \, $ from the outside, there will be 
more and more directions of $\, {\bf B} $, corresponding to the presence of 
periodic open trajectories on pairs of former carriers of stable open trajectories 
of system (\ref{MFSyst}). It can thus be stated that for direct conductivity 
measurements the ``experimentally observable'' Stability Zone 
$\, \hat{\Omega}_{\alpha} \, $ does not coincide with the exact mathematical 
Stability Zone $\, \Omega_{\alpha} \, $ and contains it as a subset 
(Fig. \ref{ExpZone}). It must be said that the analytic dependence of the 
tensor $\, \sigma^{ik} ({\bf B}) \, $ both on the direction and on the value 
of $\, {\bf B} \, $ inside the Zone $\, \hat{\Omega}_{\alpha} \, $
is generally quite complicated and can be approximated by different regimes 
in its different parts (see \cite{AnProp}). Let us also note here that the 
analytic properties of the tensor $\, \sigma^{ik} ({\bf B}) \, $ in the 
Zones $\, \hat{\Omega}_{\alpha} \, $ can play, presumably, a certain role in 
considering the conductivity of polycrystals in strong magnetic fields 
(see, e.g. \cite{DreizinDykhne1,DreizinDykhne2}), which also possesses
rather non-trivial properties in metals with complex Fermi surfaces.
At the same time, in spite of the unobservability of the exact boundaries 
of the Stability Zones in direct conductivity measurements, it is nevertheless 
possible to indicate other experimental methods that allow one to determine 
the exact mathematical Stability Zones for real materials. In particular, 
as can be shown, this problem can be solved using the study of classical or 
quantum oscillation phenomena (cyclotron resonance, de Haas - van Alphen 
effect, Shubnikov - de Haas effect, etc.) in strong magnetic fields 
(see \cite{CyclRes}).

\begin{figure}[t]
\begin{center}
\includegraphics[width=0.9\linewidth]{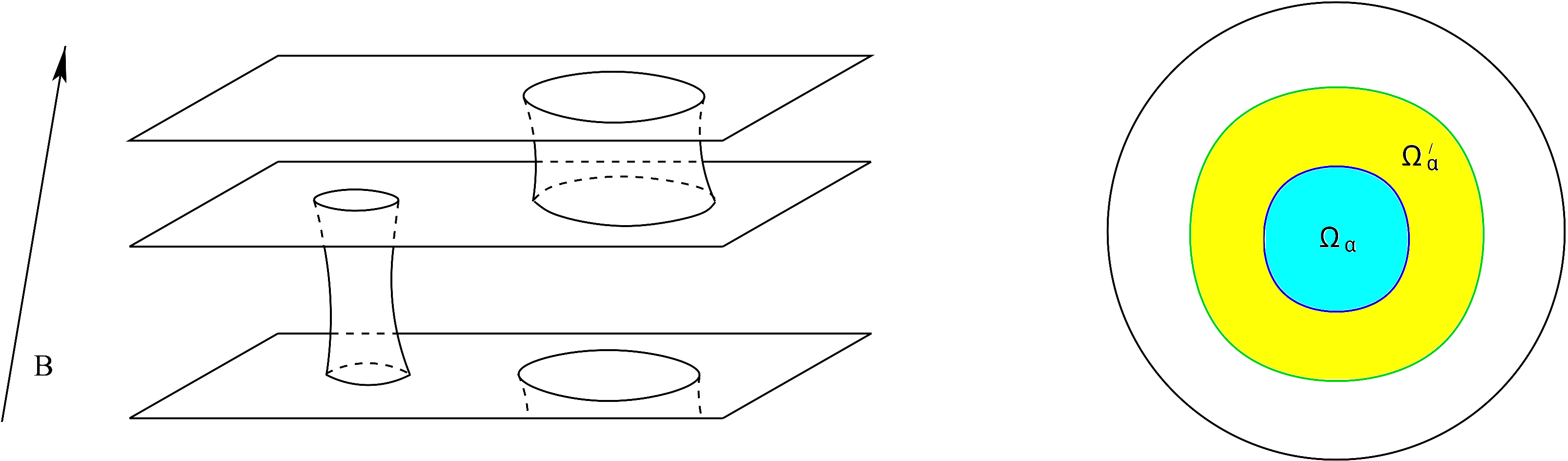} 
\end{center}
\begin{center}
\includegraphics[width=0.9\linewidth]{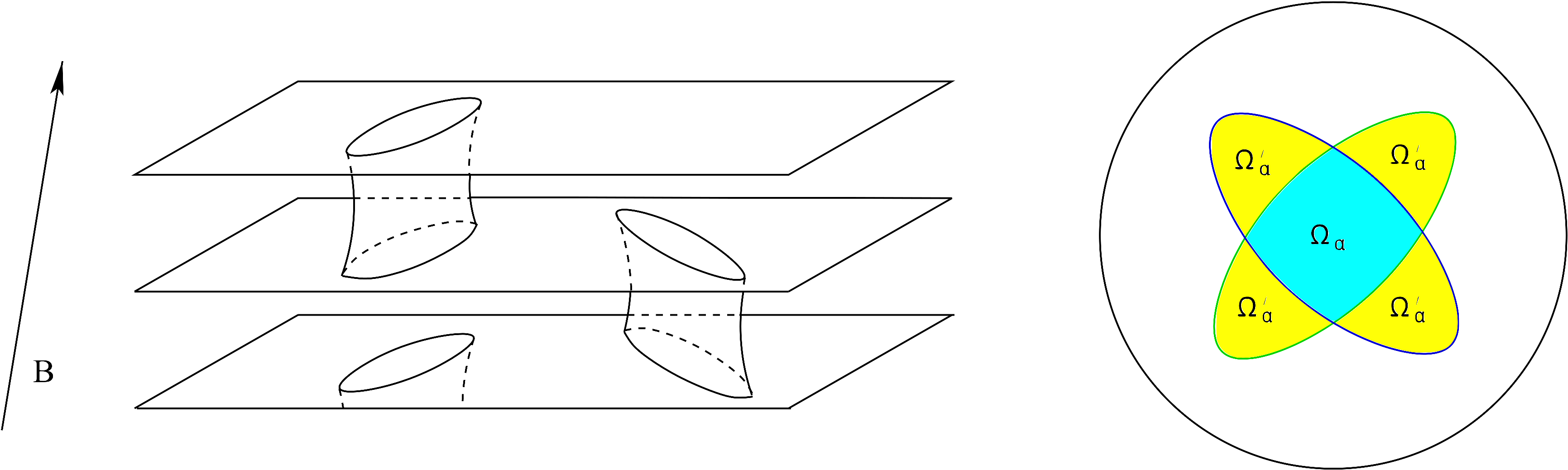} 
\end{center}
\begin{center}
\includegraphics[width=0.9\linewidth]{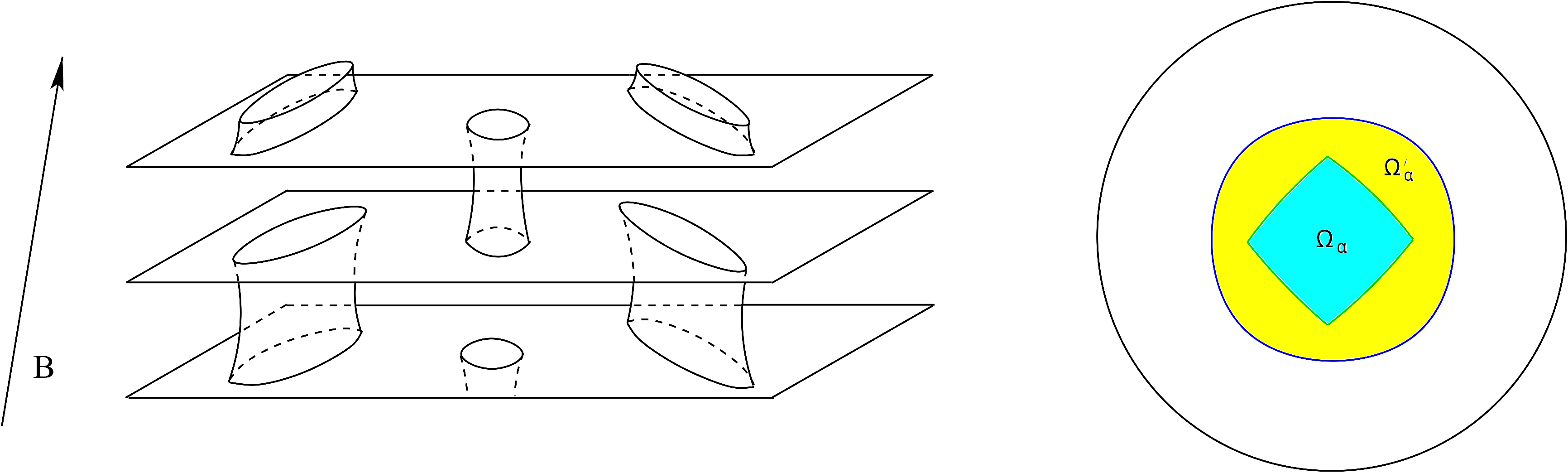}
\end{center}
\caption{The second boundaries of the Stability Zones, bounding the domains
$\, \Omega^{\prime}_{\alpha} \, $.}
\label{SecBounds}
\end{figure}

  As also follows from the arguments of the previous paragraph, each 
Stability Zone $\, \Omega_{\alpha} \, $ is in fact surrounded by some 
additional domain $\, \Omega^{\prime}_{\alpha} \, $, restricted by some 
``second boundary'' of the Stability Zone, determined by the disappearance 
of at least one more cylinder of closed trajectories presented at 
Fig. \ref{StableTrFermiSurf}. Everywhere in the region 
$\, \Omega^{\prime}_{\alpha} \, $ the Fermi surface can be represented 
as a union of pairs of former carriers of stable open trajectories
admitting jumps of the trajectories of the system (\ref{MFSyst}), 
separated by the remaining cylinders of closed trajectories.
It is easy to see that in the most common case we are considering 
(exactly two carriers of open trajectories in the Zone $\, \Omega_{\alpha} $), 
in the domain $\, \Omega^{\prime}_{\alpha} \, $ there can be no open trajectories 
other than (unstable) periodic trajectories whose direction is determined by the 
same rule as the mean direction of the open trajectories in the Zone 
$\, \Omega_{\alpha} $. It must be said that for Fermi surfaces of very 
high genus (more than two carriers of open trajectories in the Zone 
$\, \Omega_{\alpha} $), of course, we can have the situation when 
the ``merged'' carriers of open trajectories coexist with unbroken 
carriers having the same direction in $\, {\bf p}$ - space.
In this case, one can speak of a Stability Zone with a complex boundary 
or of the imposition of Stability Zones corresponding to the same topological 
quantum numbers. The region $\, \Omega^{\prime}_{\alpha} \, $ can be called 
the ``derivative'' of the Stability Zone $\, \Omega_{\alpha} $, since the 
structure of the system (\ref{MFSyst}) in this region is closely related 
to the structure of the same system in the Zone $\, \Omega_{\alpha} $.
It is easy to see that, like the first boundary of the Stability Zone, 
its second boundary can also be either simple or compound in the sense 
indicated above (Fig. \ref{SecBounds}). It can also be seen that for any 
connected part of the region $\, \Omega^{\prime}_{\alpha} \, $ its inner 
(bordering $\, \Omega_{\alpha} $) and outer boundaries correspond to the 
disappearance of the cylinders of closed trajectories of the opposite 
(electron and hole) types at Fig. \ref{StableTrFermiSurf}. We also note 
here that for a precise determination of the second boundaries of the 
Stability Zones in real crystals, we can also propose a number of experimental 
methods, including the study of classical or quantum oscillations in strong 
magnetic fields mentioned earlier (see \cite{SecBound}).

 It should also be noted that the type of cylinder of closed trajectories 
disappearing at any of the sections of the (first) boundary of a Stability Zone 
is also observable experimentally and, moreover, plays an important role in the 
conductivity behavior outside of $\, \Omega_{\alpha} $. Namely, it can be shown 
that the type of the vanishing cylinder determines the behavior of the Hall 
conductivity in strong magnetic fields in that part of the region 
$\, \Omega^{\prime}_{\alpha} \, $, where the condition 
$\, \tau \, \gg \, T \, $  starts to be fulfilled. Let us note at once that 
the behavior of the Hall conductivity in this limit is usually associated with 
the concentration of electron and hole carriers in a conductor,  
however, the type of current carriers is not determined in the presence of open 
trajectories of the system (\ref{MFSyst}) on the Fermi surface. Let us consider 
here for simplicity (the most realistic) the case when the connected component 
of the Fermi surface carrying stable open trajectories has exactly two 
nonequivalent carriers of open trajectories. In this case it can be shown 
(see \cite{SecBound}) that in calculating the Hall conductivity in the 
indicated part of $\, \Omega^{\prime}_{\alpha} \, $ one can assume that such 
a component is bounding either the carriers of the electron type, or carriers 
of the hole type, depending on the type of the vanishing cylinder of closed 
trajectories on the corresponding part of the boundary of $\, \Omega_{\alpha} $.
More precisely, the connected component must be regarded as bounding carriers 
of the electron type if on the corresponding part of the boundary of
$\, \Omega^{\prime}_{\alpha} \, $ and $ \, \Omega_{\alpha} \, $
we observe disappearance of a cylinder of closed hole-type trajectories, 
and bounding carriers of the hole type in the opposite case. To calculate 
the contribution of such a component to the Hall conductivity in the limit 
$\, \tau \, \gg \, T \, $, one can use one of the formulas
\begin{equation}
\label{Sigma12Elekt}
\sigma^{12} \,\,\, = \,\,\, 
{2 ec \over (2\pi \hbar)^{3} B} \,\,\,  V_{-}
\quad \quad {\rm (\text{the first case})}
\end{equation}
\begin{equation}
\label{Sigma12Hole}
\sigma^{12} \,\,\, = \,\,\, 
- \, {2 ec \over (2\pi \hbar)^{3} B} \,\,\,  V_{+}
\quad \quad {\rm (\text{the second case})}
\end{equation}
where $\, V_{-} \, $ and $\, V_{+} \, $ represent the volumes of the regions 
bounded by the component under consideration (in the Brillouin zone) and 
determined by conditions
$\, \epsilon ({\bf p}) \, < \, \epsilon_{F} \, $ and
$\, \epsilon ({\bf p}) \, > \, \epsilon_{F} \, $
respectively. (To calculate the total Hall conductivity, it is necessary 
to carry out summation over all connected components of the Fermi surface).

 Let us recall here that the Hall conductivity represents a ``transverse'' 
conductivity in the plane orthogonal to the magnetic field, and traditionally 
has a positive sign for positively charged current carriers and negative for 
negatively charged ones in strong magnetic fields (electron charge $\, e \, $ 
is assumed to be negative here). As we have already mentioned, for conductors 
with Fermi surfaces of general form, the carrier charge must be effectively 
considered positive or negative, depending on the trajectories of the system 
(\ref{MFSyst}). The type of current carriers is usually well defined for metals 
with simple Fermi surfaces (of rank zero) that admit only closed trajectories 
of the system (\ref{MFSyst}), and does not depend in this case on the direction 
of the magnetic field. At the same time, for metals with complex Fermi surfaces, 
the carrier type is not defined for the directions of $\, {\bf B} \, $ lying 
in the Stability Zones, as mentioned above. At the same time, the type of 
carriers can be defined for directions of $\, {\bf B} \, $ lying outside any 
of the Stability Zones, when the Fermi surface contains only closed trajectories 
of the system (\ref{MFSyst}), and this property is stable with respect to 
small rotations of $\, {\bf B} \, $. (Let us note here that this fact does not 
actually mean the existence of closed trajectories of the system (\ref{MFSyst}) 
of a fixed type (electron or hole) on each of the components of the Fermi 
surface.  Nevertheless, although on any part of the Fermi surface in this case 
there can be closed trajectories of both types, it is possible to effectively 
assign to it a fixed number of carriers of a certain type (in the Brillouin zone) 
by relating the number of carriers to one of the volumes bounded by the 
corresponding component of the Fermi surface.) As follows from the formulas 
(\ref{Sigma12Elekt}), (\ref{Sigma12Hole}), the value of the Hall conductivity 
is then locally constant (for a fixed value of $\, B $) under the condition 
$\, \tau \, \gg \, T \, $. It can be seen that in this case it is natural 
to divide the angular diagrams for conductivity into two classes, namely, 
to diagrams for which the carriers have the same charge (type A) everywhere 
outside the Stability Zones, and diagrams for which in different regions 
outside the Stability Zones carriers have different charge (type B).

\begin{figure}[t]
\begin{center}
\includegraphics[width=0.7\linewidth]{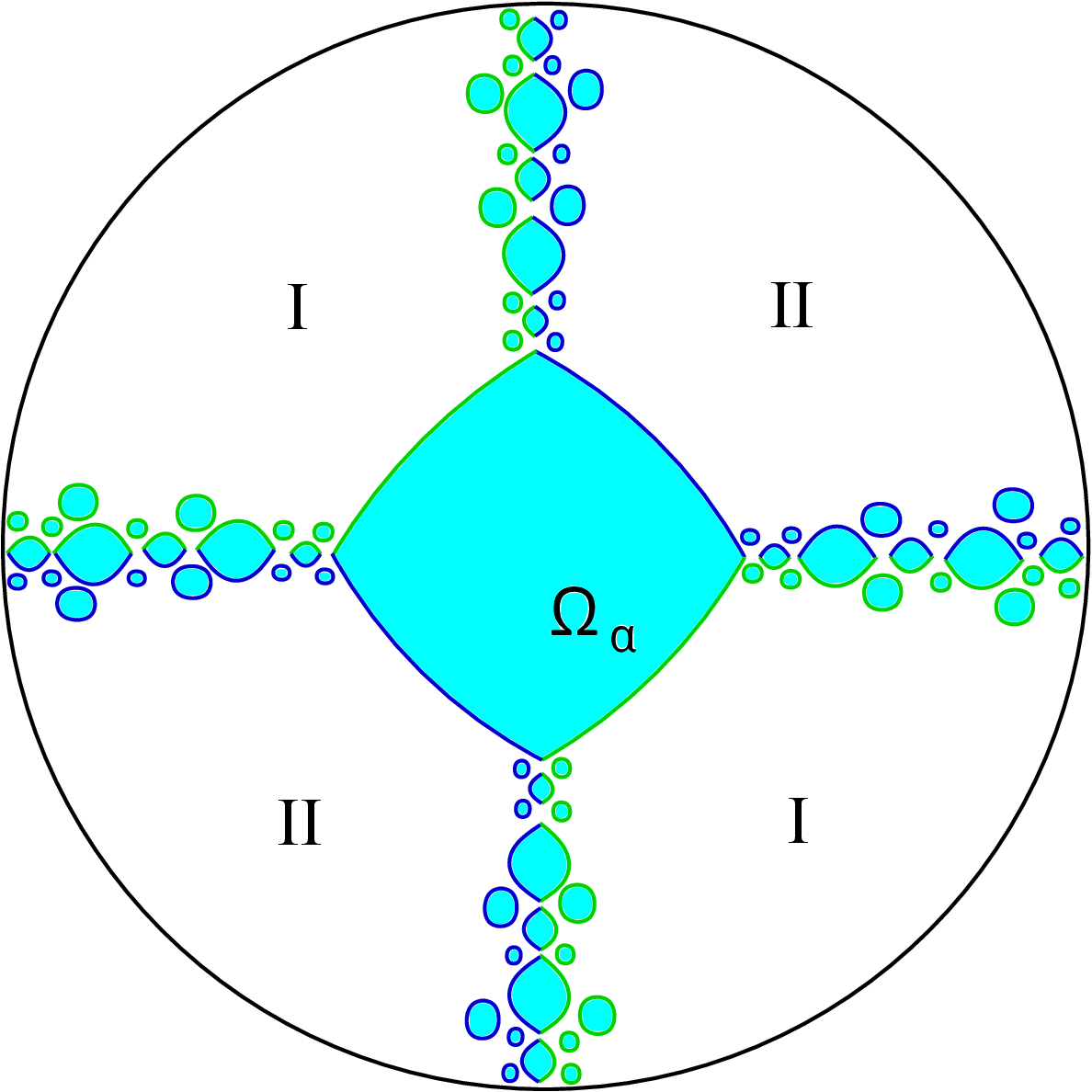}  
\end{center}
\caption{The angular diagram of type B and the Stability Zones 
separating regions corresponding to different values of the Hall 
conductivity on $\, \mathbb{S}^{2} \, $ (schematically, only the 
mathematical boundaries of a finite number of exact Stability Zones 
are shown).}
\label{TypeB}
\end{figure}

 We must at once say that diagrams of type A are a priori simpler and, 
moreover, appear, apparently, in most of the cases when studying the 
conductivity of real crystalline conductors. In particular, such diagrams 
include all diagrams containing only a finite (or infinite) number of 
Stability Zones that do not divide the sphere $\, \mathbb{S}^{2} \, $ into 
unrelated domains. As for the B-type diagrams, it is easy to see that the 
Stability Zones should form a rather complex structure here, splitting 
$\, \mathbb{S}^{2} \, $ into regions with different types of current carriers.
Nevertheless, from a theoretical point of view, B-type diagrams also represent 
general diagrams, and in particular arise each time when there is at least one 
Stability Zone with a compound boundary defined by the disappearance of 
cylinders of closed trajectories of different types on its different parts
(Fig. \ref{CompoundFirstBound}). As already noted above, in this case we 
necessarily have a situation when in different parts of the region 
$\, \Omega^{\prime}_{\alpha} \, $ the corresponding component of the 
Fermi surface corresponds to carriers of different types for generic 
directions of $\, {\bf B} $. It can also be noted that the type of 
carriers does not change in generic case also after crossing 
the second boundary of a Stability Zone (see \cite{SecBound}), so that 
the regions of different types of charge carriers that arise in this case 
have a rather complex structure, separated by ``chains'' of Stability Zones 
on $\, \mathbb{S}^{2} \, $ (Fig. \ref{TypeB}). Note also that the ``chains'' 
contain generically an infinite number of Stability Zones. Thus, 
for example, the ``corner'' point of the Stability Zone at 
Fig. \ref{CompoundFirstBound} can be adjoined by another Stability Zone 
only if the corresponding direction of $\, {\bf B} \, $ is associated with 
the appearance of periodic open trajectories on the Fermi surface.
In the case of a general position the corner point of the Zone at 
Fig. \ref{CompoundFirstBound} must be adjoined by a chain of an infinite 
number of decreasing Stability Zones.

 As we have already said, diagrams of type A are, apparently, the main type 
that arises for real crystal conductors. At the same time, the detection of 
a B-type diagram for some real material would allow us to observe a wide 
variety of different behavior regimes for the magnetoconductivity for different 
directions of $\, {\bf B} \, $. It can be noted at the same time that despite 
the possible experimental difficulties in observing the entire complex picture 
of the Stability Zones arising in this case, the identification of the type of 
the diagram can be easily established by the behavior of the Hall conductivity 
outside the (experimentally observable) Stability Zones. We note here that,
since the total Fermi surface can consist of several components that contribute 
to the Hall conductivity, a B-type diagram differs in the general case by the 
presence of at least two different values of Hall conductivity 
(no matter what sign) in different domains outside the experimentally 
observable Stability Zones at a given value of $\, B $.

 We can also define extended Stability Zones
$$ \Sigma_{\alpha} \,\,\, = \,\,\, \Omega_{\alpha} \, \cup \, 
\Omega^{\prime}_{\alpha} \quad  ,  $$
defined by the union of domains $\, \Omega_{\alpha} \, $ and
$\, \Omega^{\prime}_{\alpha} \, $. It is not difficult to see then that 
the open trajectories of the system (\ref{MFSyst}), other than those considered 
above, can arise only on the set
$$ \mathbb{S}^{2} \, \setminus \, \cup_{\alpha} \bar{\Sigma}_{\alpha} $$
of directions of $\, {\bf B} \, $ on the angular diagram. We note here that 
the size and shape of the extended Stability Zones are not related in general 
to the size of the experimentally observable Stability Zones 
$\, \hat{\Omega}_{\alpha} \, $, so that the Zones $\, \Sigma_{\alpha} \, $
can be both subsets of $\, \hat{\Omega}_{\alpha} \, $, or contain 
$\, \hat{\Omega}_{\alpha} \, $ as subsets. Note also that, unlike Zones 
$\, \Omega_{\alpha} \, $, the Zones $\, \Sigma_{\alpha} \, $ can overlap 
with each other. In particular, if the Zones $\, \Sigma_{\alpha} \, $ cover 
the entire sphere $\, \mathbb{S}^{2} \, $, this means that on the given Fermi 
surface only stable or periodic open trajectories of the system (\ref{MFSyst})
can appear. 

 Thus, in the most general case, on the angular diagram of conductivity we can 
have a finite or infinite number of disjoint mathematical Stability Zones 
$\, \Omega_{\alpha} \, $ corresponding to different topological numbers 
$\, (M^{\alpha}_{1}, M^{\alpha}_{2}, M^{\alpha}_{3}) \, $
and covering some part of the sphere $\, \mathbb{S}^{2} \, $.
On the remaining part of $\, \mathbb{S}^{2} \, $ for almost all directions 
of $\, {\bf B} \, $ the Fermi surface contains only closed trajectories of 
the system (\ref{MFSyst}). For special directions of $\, {\bf B} \, $, 
nevertheless, unstable periodic trajectories of the system (\ref{MFSyst}) 
or chaotic trajectories of Tsarev or Dynnikov type (they will be considered 
in the next chapter) may appear on the Fermi surface. As we have already said, 
theoretically, all the angular diagrams of conductivity can be divided into 
two types A and B. The diagrams of the second type differ in this case by the 
presence of regions outside the Stability Zones with different values of 
the Hall conductivity in strong magnetic fields and an infinite number of 
Stability Zones $\, \Omega_{\alpha} \, $ (generically) which divide these 
regions in the general case. As we have already noted, in the experiments 
on direct measurement of the magnetoconductivity, the exact mathematical 
Stability Zones are usually included in the wider ``experimentally observable'' 
Stability Zones $\, \hat{\Omega}_{\alpha} \, $. In addition, as was also noted 
above, for each Stability Zone $\, \Omega_{\alpha} \, $ we can indicate 
an adjoining domain $\, \Omega^{\prime}_{\alpha} \, $, where the description 
of trajectories of the system (\ref{MFSyst}) is actually closely related 
to the structure of (\ref{MFSyst}) in the Zone $\, \Omega_{\alpha} \, $
and does not allow the appearance of open trajectories of (\ref{MFSyst}), 
other than periodic. We note here again that the described structure of 
the angular diagram of conductivity holds for all Fermi surfaces 
(not necessarily defined by a single dispersion relation) consisting of 
components that do not intersect each other.

\vspace{1mm}

 Returning to the general problem of describing the geometry of the 
trajectories of the system (\ref{MFSyst}) with an arbitrary dispersion law, 
it is also necessary to give a description of the angular diagrams for 
the total dispersion relation $\, \epsilon ({\bf p}) \, $, introduced 
by I.A. Dynnikov in the work \cite{dynn3}. Let us formulate here the main 
statements given in \cite{dynn3}, which represent a basis for describing 
such diagrams.

\vspace{1mm}

 Consider an arbitrary dispersion relation given by a smooth 3-periodic 
function $\, \epsilon ({\bf p}) \, $, such that
$\, \epsilon_{min} \, \leq \, \epsilon ({\bf p}) \, \leq \,
\epsilon_{max} \, $. Let us fix an arbitrary direction of $\, {\bf B} \, $ 
and consider the energy levels $\, \epsilon ({\bf p}) \, = \, const \, $
containing open trajectories of the system (\ref{MFSyst}) in the 
extended $\, {\bf p}$ - space.

 Then:

\vspace{1mm}
 
 Energy levels containing open trajectories of (\ref{MFSyst}) represent 
either a connected interval
$$\epsilon_{1} ({\bf B}/B) \,\,\, \leq \,\,\, \epsilon
\,\,\, \leq \,\,\, \epsilon_{2} ({\bf B}/B) $$
or only one isolated point
$\, \epsilon \, = \, \epsilon_{0} ({\bf B}/B) $.

\vspace{1mm}

 For generic directions of $\, {\bf B} \, $ the boundaries of the 
interval of open trajectories $\, \epsilon_{1} ({\bf B}/B) \, $ and
$\, \epsilon_{2} ({\bf B}/B) \, $ are determined by the values 
of some globally defined continuous functions
$\, \tilde{\epsilon}_{1} ({\bf B}/B) \, $ and
$\, \tilde{\epsilon}_{2} ({\bf B}/B) \, $.
At the same time, for special directions of $\, {\bf B} \, $, corresponding 
to appearance of periodic  open trajectories of system (\ref{MFSyst}),  we can 
write the relations
\begin{equation}
\label{SpecialRel}
\epsilon_{1} ({\bf B}/B) \,\,\, \leq \,\,\, 
\tilde{\epsilon}_{1} ({\bf B}/B) \,\,\, , \quad 
\epsilon_{2} ({\bf B}/B) \,\,\, \geq \,\,\, 
\tilde{\epsilon}_{2} ({\bf B}/B)
\end{equation}

\vspace{1mm}

 Every time when we have the situation
$\,  \tilde{\epsilon}_{2} ({\bf B}/B) \, > \, 
\tilde{\epsilon}_{1} ({\bf B}/B) \, $,
all the (non-singular) open trajectories of system (\ref{MFSyst}),
arising for generic directions of $\, {\bf B} \, $, 
have a regular shape, represented at Fig. \ref{StableTr}, and the same 
mean direction given by the intersection of the plane orthogonal to 
$\, {\bf B} \, $ and some integral plane $\, \Gamma \, $ 
in the $\, {\bf p}$ - space.

\begin{figure}[t]
\begin{center}
\includegraphics[width=0.7\linewidth]{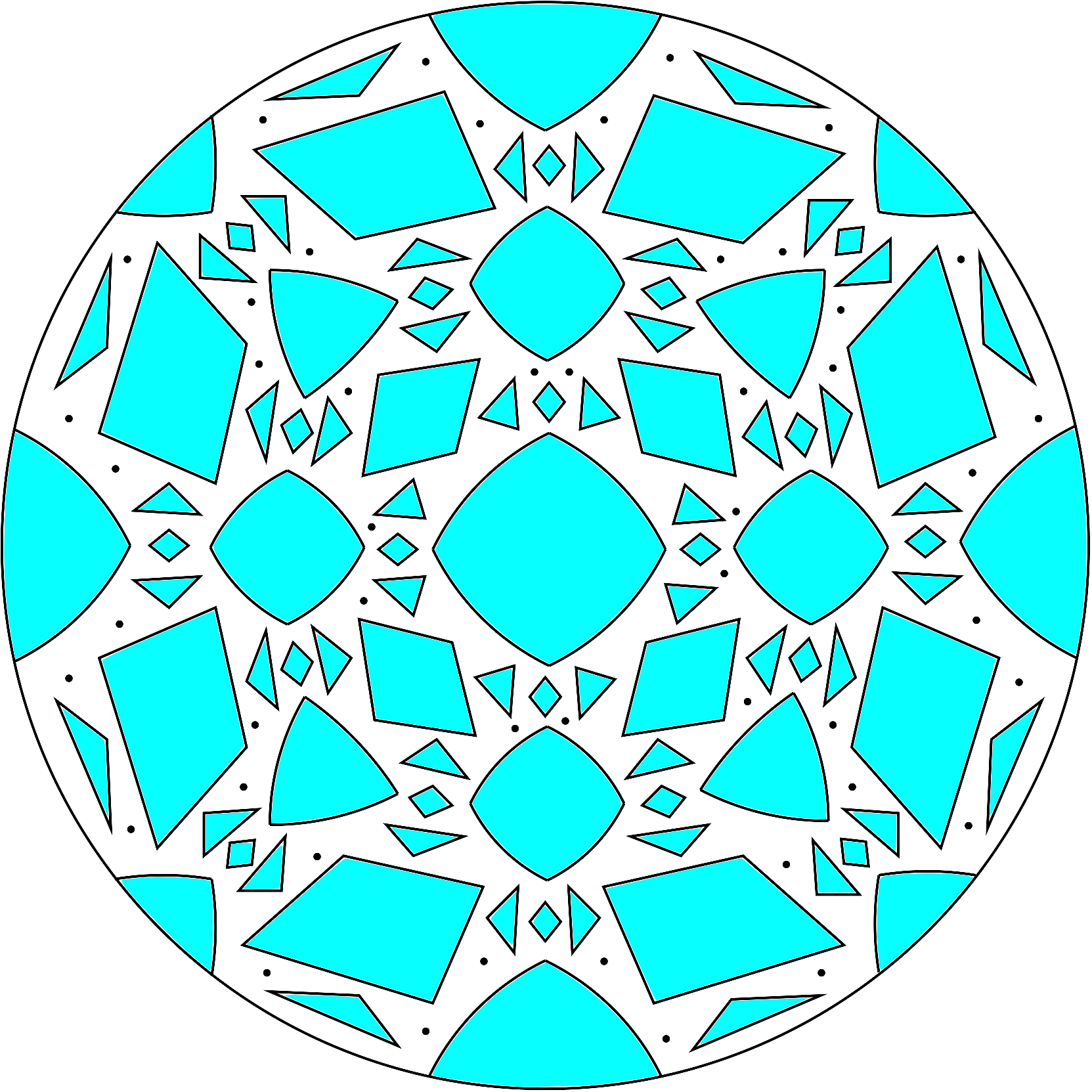}  
\end{center}
\caption{The angular diagram for a general dispersion relation 
$\, \epsilon ({\bf p}) \, $ (schematically, only a finite number 
of decreasing Stability Zones $\, \Omega^{*}_{\alpha} \, $ and 
special ``chaotic'' directions is shown).}
\label{DispRel}
\end{figure}

 All such trajectories, arising for generic directions of $\, {\bf B} \, $,
and also the integral plane $\, \Gamma \, $ are stable with respect to
small rotations of $\, {\bf B} \, $, and the complete set of directions 
of $\, {\bf B} \, $, corresponding to the same plane $\, \Gamma \, $,
represents a finite domain with piecewise smooth boundary on the angular 
diagram $\, \mathbb{S}^{2}$. The regions $\, \Omega^{*}_{\alpha} \, $, 
corresponding to the presence of stable open trajectories with the same 
integral plane $\, \Gamma_{\alpha} \, $ on any of the energy levels, 
represent in this case the Stability Zones for the entire dispersion 
relation $\, \epsilon ({\bf p}) \, $. On the boundaries of the Zones
$\, \Omega^{*}_{\alpha} \, $ we always have the relation
$\, \tilde{\epsilon}_{1} ({\bf B}/B) \, = \,
\tilde{\epsilon}_{2} ({\bf B}/B) \, $. Let us also note here that the 
boundaries of the Stability Zones $\, \Omega^{*}_{\alpha} \, $ have 
in this case both the ``electron'' and the ``hole'' type, and are 
determined by the disappearance of simultaneously two cylinders of closed 
trajectories of opposite types when crossing them. The complete set 
of Stability Zones $\, \Omega^{*}_{\alpha} \, $ forms an everywhere 
dense set on the angular diagram (Fig. \ref{DispRel}), which can generally 
contain either one or infinitely many different Stability Zones.

\vspace{1mm}

 Let us also note that on the boundaries of the Zones 
$\, \Omega^{*}_{\alpha} \, $ we have open trajectories of the system 
(\ref{MFSyst}) having a regular form, shown at Fig. \ref{StableTr}, 
which in this case are not stable with respect to all small rotations 
of $\, {\bf B} \, $, as well as energy level variations.

\vspace{1mm}

 It is not difficult to see that the angular diagram of a Fermi surface, 
determined by one dispersion relation
$\, \epsilon ({\bf p}) = \epsilon_{F} \, $,
can be ``nested'' in the diagram of the total dispersion relation, 
so that the Stability Zones, defined for a fixed Fermi surface, represent 
subsets of a part of the Zones $\, \Omega^{*}_{\alpha} \, $. For Fermi surfaces 
defined by several dispersion relations, in general, such a natural embedding 
can be absent. It should be noted here that the latter situation arises actually 
only for sufficiently complex Fermi surfaces containing several complex components.

\vspace{1mm}

 The complement to the set 
$\, \left\{ \cup \, \overline{\Omega^{*}_{\alpha}} \right\} \, $ on the sphere 
$\, \mathbb{S}^{2} \, $ forms a rather complex set (of Cantor type) and represents 
the directions of $\, {\bf B} \, $ corresponding to the appearance of chaotic open
trajectories at one energy level
$$\epsilon_{0} ({\bf B}/B) \,\,\, = \,\,\, 
\tilde{\epsilon}_{1} ({\bf B}/B) \,\,\, = \,\,\,
\tilde{\epsilon}_{2} ({\bf B}/B) $$

 According to the conjecture of S.P. Novikov (see \cite{DynSyst}), this set 
has measure zero and the Hausdorff dimension strictly less than 2 on the unit 
sphere. We can note here that the structure of the set of such special directions 
and the properties of chaotic trajectories are actively investigated at the present 
time (see, e.g. \cite{dynn2,dynn3,ZhETF2,zorich2,
Zorich1996, ZorichAMS1997,zorich3,DeLeo1,DeLeo2,DeLeo3,ZorichLesHouches,
DeLeoDynnikov1,dynn4,DeLeoDynnikov2,Skripchenko1,Skripchenko2,DynnSkrip1,
DynnSkrip2,AvilaHubSkrip1,AvilaHubSkrip2,DeLeo2017}).

\vspace{1mm}

 In conclusion of this chapter, it can be noted that although angular diagrams 
of conductivity for the complete dispersion relation are not yet observed in the 
experiment, it is possible that they will still be observed in the measurement of 
(photoinduced) conductivity in semiconductors in extremely strong magnetic fields 
(see \cite{JETP1}).

\section{Chaotic trajectories and two-dimensional electron systems.}
\setcounter{equation}{0}

\begin{figure}[t]
\begin{tabular}{ll}
\includegraphics[width=0.45\linewidth]{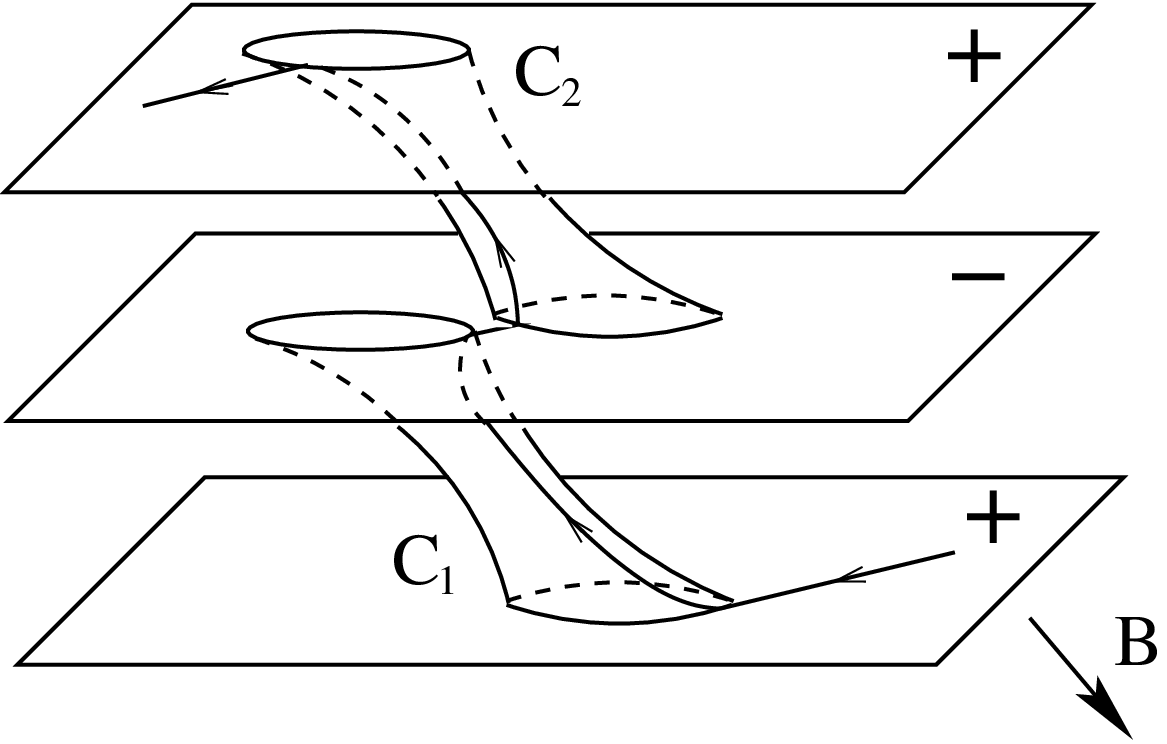}  &
\hspace{5mm}
\includegraphics[width=0.45\linewidth]{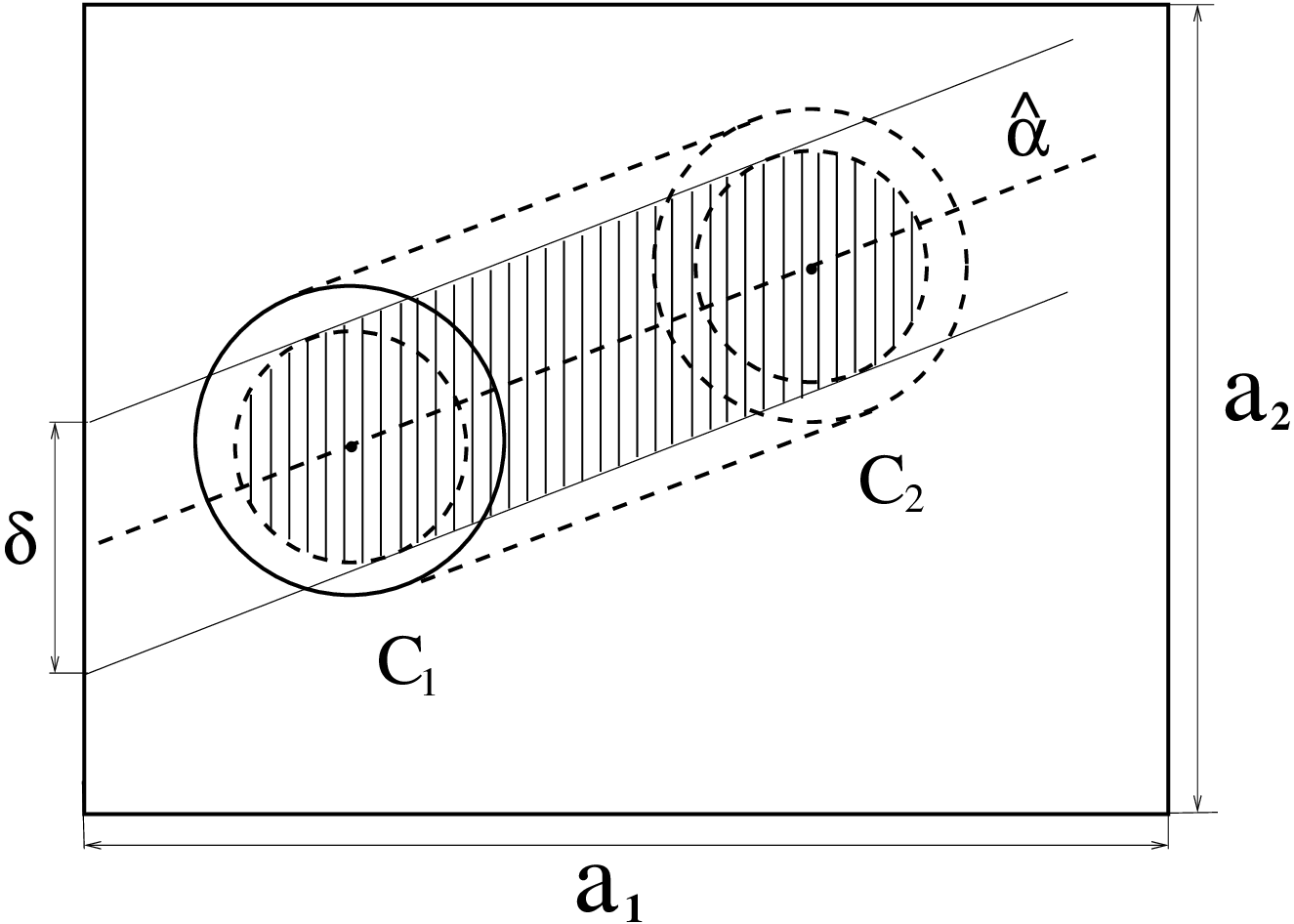} 
\end{tabular}
\caption{The Brillouin zone and the behavior of the trajectories of the 
system (\ref{MFSyst}) for a Fermi surface carrying chaotic trajectories 
of Tsarev type.}
\label{TsarevExample}
\end{figure}

 In this chapter we will consider the trajectories of the system (\ref{MFSyst}), 
which have chaotic properties. We will start with a simpler example, constructed 
by S.P. Tsarev (private communication, 1992-1993). The general idea of constructing
chaotic trajectories, suggested by Tsarev, can be expressed by the following scheme:

 Consider a family of identical integral (horizontal) planes connected 
by identical cylinders, as shown at Fig. \ref{TsarevExample}. Let us assume 
that the centers of all the bases of the cylinders shown at 
Fig. \ref{TsarevExample} lie in one (for simplicity, vertical) plane 
intersecting the horizontal planes in some direction $\, \hat{\alpha} $.
We also assume that the constructed cylinders are repeated periodically, 
so that the constructed surface is periodic with some periods $\, {\bf a}_{1} $,
$\, {\bf a}_{2} $, lying in the horizontal plane, and also a vertical period 
$\, {\bf a}_{3} $. It is not difficult to see that the constructed surface 
can be regarded as a periodic Fermi surface, and we must divide all the 
integral planes (even and odd), and the cylinders ($C_{1}$ and $C_{2}$) 
into two different classes. Let us consider a horizontal magnetic field 
$\, {\bf B} $, orthogonal to the direction $\, \hat{\alpha} $, and the 
trajectories of the system (\ref{MFSyst}) corresponding to this direction. 
The trajectories of the system (\ref{MFSyst}) can be considered as trajectories 
on integral planes (of two different types), sometimes jumping from one plane 
to another. Under the above conditions, it is not difficult to see that any 
trajectory that jumps from a plane to an adjacent plane inevitably jumps 
to the next plane and continues in the same direction as on the original plane 
(Fig. \ref{TsarevExample}). It is easy to see here that all the (non-singular) 
trajectories of the system (\ref{MFSyst}) have an asymptotic direction in 
the $\, {\bf p}$ - space defined by the relations between the effective radius 
of the cylinders and the periods $\, {\bf a}_{1} $, $\, {\bf a}_{2} $,
$\, {\bf a}_{3} $. At the same time, for any irrational direction 
$\, \hat{\alpha} \, $ (with respect to the given crystal lattice), no regular 
trajectory of (\ref{MFSyst}) can lie in a straight line of finite width in the 
corresponding plane, orthogonal to $\, {\bf B} $.

 It is easy to see, that in the above construction each regular trajectory 
of the system (\ref{MFSyst}) sweeps out half the surface of genus 3 and, 
in this sense, has a chaotic behavior on the Fermi surface. In the extended 
$\, {\bf p}$ - space, however, the trajectories of Tsarev are more like 
the trajectories described in the previous chapter, having asymptotic 
directions in the plane orthogonal to $\, {\bf B} $. As was shown in 
\cite{dynn2}, the last property is actually observed for all chaotic 
trajectories arising for directions of $\, {\bf B} \, $  of irrationality 2 
(the plane orthogonal to $\, {\bf B} $ contains a reciprocal lattice vector).
As in the case of stable open trajectories, the contribution of Tsarev's 
chaotic trajectories to magnetoconductivity has a more complex analytic 
behavior than that given by the formula (\ref{Periodic}), but has similar 
geometric properties. In particular, here also the general formula 
(\ref{GeneralLimit}) takes place with a proper choice of the coordinate 
system.

\begin{figure}[t]
\begin{center}
\includegraphics[width=\linewidth]{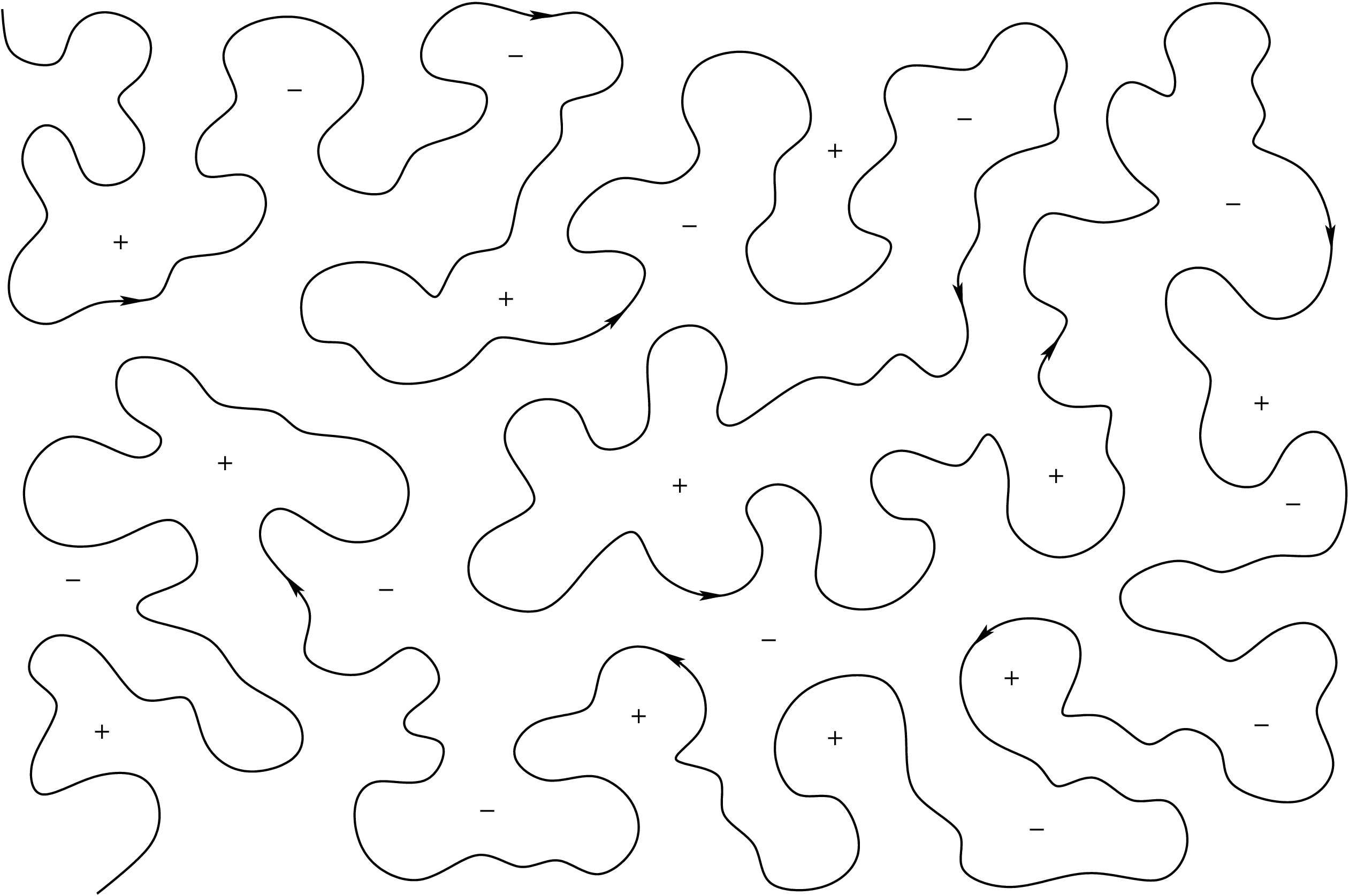} 
\end{center}
\caption{Dynnikov chaotic trajectory in a plane orthogonal to 
$\, {\bf B} $.}
\label{DynnChaoticTr}
\end{figure}

 More complex examples of chaotic trajectories of the system (\ref{MFSyst}) 
were first constructed by I.A. Dynnikov in the work \cite{dynn2}. The Dynnikov 
trajectories arise for directions of $\, {\bf B} \, $ of maximal irrationality 
and have complex chaotic behavior both on the Fermi surface and in the extended 
$\, {\bf p}$ - space. As a rule, the Dynnikov trajectories everywhere densely 
sweep components of genus 3 (or more) with contractible holes in the Brillouin 
zone, having an obvious chaotic behavior on these components. In the extended 
$\, {\bf p}$ - space, the behavior of such trajectories in planes orthogonal 
to $\, {\bf B} $, resembles diffusional motion to some extent, although, 
of course, it is not diffusion in the strict sense of the word 
(Fig. \ref{DynnChaoticTr}). In general, the behavior of trajectories of this 
type on the Fermi surface and in the extended $\, {\bf p}$ - space represents 
an important example of an emergence of classical chaos in the condensed 
matter physics.

 The behavior of the magnetoconductivity (and other transport phenomena) 
in strong magnetic fields in the presence of chaotic trajectories of Dynnikov 
type on the Fermi surface has the most complicated form. The most interesting 
phenomenon arising in this case is the suppression of the conductivity along 
the direction of the magnetic field in the limit
$\, \omega_{B} \tau \rightarrow \infty \, $ (\cite{ZhETF2}).
Thus, because the chaotic trajectory sweeps out a part of the Fermi surface, 
which is invariant under the transformation 
$\, {\bf p} \, \rightarrow \, - {\bf p} \, $,
the contribution of this part to the longitudinal conductivity disappears 
in the above limit. In this situation, the longitudinal conductivity is created 
only by the remaining part of the Fermi surface filled with closed trajectories 
of the system (\ref{MFSyst}). As a consequence of this fact, the longitudinal 
conductivity must have sharp minima for the directions of $\, {\bf B} $ 
corresponding to the appearance of such trajectories on the Fermi surface.

 Another interesting circumstance arising in the description of transport 
phenomena in the presence of Dynnikov-type trajectories on the Fermi surface 
is the appearance of fractional powers of the parameter $\, \omega_{B} \tau \, $
in the asymptotic behavior of the components of the tensor 
$\, \sigma^{ik} (B) \, $ as $\, \omega_{B} \tau \rightarrow \infty \, $
(\cite{ZhETF2,TrMian}). Let us note here that the first indication of this 
circumstance in \cite{ZhETF2} was actually based on some additional property 
of the trajectories (self-similarity) constructed in \cite{dynn2}. It must be 
said that this property is not, generally speaking, common for trajectories 
of the Dynnikov type. Nevertheless, as can be shown (see \cite{TrMian}), 
the appearance of such powers is in fact a more general fact and is associated 
with important characteristics (the Zorich - Kontsevich - Forni indices) 
of dynamical systems on surfaces. The Zorich - Kontsevich - Forni indices 
play an important role in describing the behavior of the trajectories of 
dynamical systems and can be determined for a fairly wide class of dynamical 
systems and foliations on surfaces 
(see \cite{zorich2,Zorich1996,ZorichAMS1997,Zorich1997,zorich3,ZorichLesHouches}).
Let us also note here that for the dynamical system (\ref{MFSyst}) the existence 
of the Zorich - Kontsevich - Forni indices, strictly speaking, requires additional 
justification and does not follow automatically from the general theory based 
on certain generic requirements. As an example of such a justification, we can 
indicate the work \cite{AvilaHubSkrip1}, in which the construction and 
investigation of Dynnikov type trajectories was carried out, and the existence 
of the indicated indices for the Fermi surface of a rather general form was 
established.

 We give here a general description of the Zorich - Kontsevich - Forni indices 
defined in the general case for foliations generated by closed 1-forms on compact 
surfaces $\, M^{2}_{g} $. We will follow here the work \cite{zorich3}, where for 
``almost all'' foliations of this type the following properties were 
indicated:

 For a foliation generated by a closed 1-form on a surface of genus 
$\, g $, we consider a layer (level line) in general position and fix an initial 
point $\, P_{0} \, $ on it. On the same layer, we fix another point $\, P_{1} \, $,
in which this layer approaches close to $\, P_{0} \, $ after passing a sufficiently 
large path along the surface $\, M^{2}_{g} $. We join the points $\, P_{0} \, $ and 
$\, P_{1} \, $ by a short segment and define a closed cycle on the surface 
$\, M^{2}_{g} $. Let us denote the homology class of the resulting cycle by 
$\, c_{P_{0}} (l) \, $, where $\, l \, $ represents the length of the corresponding 
section of the layer in some metric. Then:

 There is a flag of subspaces
$$V_{1} \,\, \subset \,\, V_{2} \,\, \subseteq \,\, \dots \,\, 
\subseteq \,\, V_{g} \,\, \subseteq \,\, 
V \,\, \subset \,\, H_{1} ( M^{2}_{g}; \mathbb{R} ) \,\,\, , $$
such that:

 1) For any such layer $\, \gamma \, $ and any point
$\,\, P_{0} \, \in \, \gamma \,\, $  
$$\lim_{l \rightarrow \infty} \,\, {c_{P_{0}} (l) \over l} \,\,\, = \,\,\, c \,\,\, , $$
where the non-zero asymptotic cycle
$\,\, c \, \in \, H_{1} ( M^{2}_{g}; \mathbb{R} ) \,\, $  is proportional to the
Poincare cycle and generates the subspace $\, V_{1} \, $.

 2) For any linear form
$\,\,\phi \, \in \, Ann (V_{j}) \, \subset \, H_{1} ( M^{2}_{g}; \mathbb{R} ) \, $,
$\,\, \phi \notin \, Ann (V_{j+1}) \,\, $
\begin{equation}
\label{LimSupRel}
\limsup_{l \rightarrow \infty} \,\,
{\log |\langle \phi ,  c_{P_{0}} (l) \rangle | \over \log l} \,\,\, = \,\,\, \nu_{j+1}
\quad , \quad \quad j \, = \, 1, \, \dots , \, g -1 
\end{equation}

3) For any
$\,\,\phi \, \in \, Ann (V) \, \subset \, H_{1} ( M^{2}_{g}; \mathbb{R} ) \, $,
$\,\, || \phi || \, = \, 1 \,\, $  
$$|\langle \phi ,  c_{P_{0}} (l) \rangle | \,\, \leq \,\, const  \,\,\, , $$
where the constant is determined only by foliation.

 4) The subspace $\,\, V \, \subset \, H_{1} ( M^{2}_{g}; \mathbb{R} ) \,\, $
is Lagrangian in homology, where the symplectic structure is determined by the 
intersection form.

 5) Convergence to all the above limits is uniform in 
$\, \gamma \, $ and $\,\, P_{0} \, \in \, \gamma \, $, i.e. depends only on
$\, l \, $.

\vspace{1mm}

 It can be seen that the above statements give extremely important information 
about the behavior of level lines of a foliation (or trajectories of a dynamical 
system) on the manifold $\, M^{2}_{g} $. It can also be shown that the properties 
described above also significantly affect the behavior of chaotic trajectories 
in the extended $\, {\bf p}$ - space under the condition that 
the Zorich - Kontsevich - Forni indices exist for the system (\ref{MFSyst}) 
on the Fermi surface. Indeed, suppose that for some system (\ref{MFSyst}) 
chaotic Dynnikov trajectories arise on a complex Fermi surface and fill a part 
of the Fermi surface bounded by closed singular trajectories. For greater 
rigorousness, we can use the procedure of gluing the corresponding holes in 
the $\, {\bf p}$ - space and define a new smooth Fermi surface carrying chaotic 
trajectories of the same global geometry as the system (\ref{MFSyst}) 
(see \cite{dynn3}). We also put for simplicity (as is the case for the most 
realistic situations), that the corresponding carriers of chaotic trajectories 
have genus 3. 
 
 Under the condition that the Zorich - Kontsevich - Forni indices exist for 
the system under consideration, we can in this case speak of the presence of 
a flag of subspaces
$$V_{1} \,\, \subset \,\, V_{2} \,\, \subseteq \,\, V_{3} \,\, \subseteq \,\, 
V \,\, \subset \,\, H_{1} ( M^{2}_{3}; \mathbb{R} ) \,\,\, , $$
possessing the properties listed above.

 We note that in the generic case we assume that
$\,\, 1 \, > \, \nu_{2} \, > \, \nu_{3} \, > \, 0 \, $,
$\,\, dim \, V_{2} \, = \, 2 \, $,
$\,\, dim \, V_{3} \, = \, dim \, V \, = \, 3 \, $.

 To describe the properties of the trajectories of the system (\ref{MFSyst}) 
in the extended $\, {\bf p}$ - space which we need, let us now consider the 
map in homology
$$H_{1} ( M^{2}_{g}; \mathbb{R} ) \,\,\, \rightarrow \,\,\,
H_{1} ( \mathbb{T}^{3}; \mathbb{R} ) \,\,\, , $$
induced by the embedding
$\,\, M^{2}_{3} \, \subset \, \mathbb{T}^{3} \, $.
It is not difficult to see that the images of all spaces $\, V_{j} \, $ must 
belong to a two-dimensional subspace defined by the plane orthogonal to 
$\, {\bf B} \, $. In addition, from the absence of a linear growth of 
the deviation from the point $\, P_{0} \, $ with increasing length of a 
trajectory in the plane orthogonal to $\, {\bf B} \, $ in examples of chaotic 
trajectories of Dynnikov type we get that the image of the asymptotic cycle 
$\, c \, $ is equal to zero under this mapping. The image of the subspace 
$\, V_{2} \, $ in the generic case is one-dimensional and determines the 
selected direction in the plane orthogonal to $\, {\bf B} \, $, along which 
the average deviation of the trajectory grows faster with its length 
$\,( \sim l^{\nu_{2}}) \, $, than in the direction orthogonal to it.
In the general case, we must also assume that the image of the space 
$\, V_{3} \, $ is two-dimensional and coincides with the plane orthogonal 
to $\, {\bf B} \, $. Considering the 1-forms $\, d p_{x} \, $ and 
$ \, d p_{y} \, $ as a basis of 1-forms, subject to the above condition (2),
we can in our situation choose the coordinate system 
$\, (x, y, z) \, $ in such a way that for some reference sequences of 
values $ \, l \, $ we will have the relations
$$| \Delta p_{x} (l) | \,\,\, \simeq \,\,\, p_{F} \, 
\left( {l \over p_{F}} \right)^{\nu_{2}} \quad , \quad \quad
| \Delta p_{y} (l) | \,\,\, \simeq \,\,\, p_{F} \,
\left( {l \over p_{F}} \right)^{\nu_{3}} $$
for the deviations of the trajectory along the coordinates 
$\, p_{x} \, $ and $\, p_{y} \, $ when passing a part of the approximate 
cycles on the Fermi surface described above. 

 It can thus be seen that the existence of the Zorich - Kontsevich - Forni 
indices predetermines certain properties of ``wandering'' of electron 
trajectories in the extended $\, {\bf p}$ - space, and, consequently, 
in the coordinate space, according to the specifics of the electron 
motion in magnetic fields. In turn, as well as for closed or stable open 
trajectories, the geometric properties of chaotic trajectories have a decisive 
influence on the characteristics of electron transport phenomena in strong 
magnetic fields. In particular, we can expect here the manifestation of 
the Zorich - Kontsevich - Forni indices in the study of the magnetoconductivity 
in crystals in the limit $\, \omega_{B} \tau \rightarrow \infty \, $.

 Indeed, as a more detailed analysis of the kinetic equations in this 
situation shows (see \cite{TrMian}), the existence of the 
Zorich - Kontsevich - Forni indices is manifested directly in the behavior 
of the components of the conductivity tensor in the plane,
orthogonal to $\, {\bf B} $, and, in particular, leads to the following
dependencies of the components $\, \sigma^{xx} \, $ and $\, \sigma^{yy} \, $
on the magnetic field:
$$\sigma^{xx} (B) \,\,\, \simeq \,\,\,
{n e^{2} \tau \over m^{*}} \,
\left( \omega_{B} \tau \right)^{2\nu_{3}-2} \quad , \quad \quad
\sigma^{yy} (B) \,\,\, \simeq \,\,\,
{n e^{2} \tau \over m^{*}} \, 
\left( \omega_{B} \tau \right)^{2\nu_{2}-2} $$

 It should be noted here that the above relations do not in fact represent 
the principal term of any asymptotic expansion of the quantities 
$\, \sigma^{xx} (B) \, $ and $\, \sigma^{yy} (B) \, $ and define more likely 
some common ``trend'' in their behavior. Strictly speaking, it is also more 
correct here to write the relations
$$\limsup_{\omega_{B} \tau \rightarrow \infty} \,\, 
{\log \sigma^{xx} (B) \over \log \omega_{B} \tau} 
\,\,\, = \,\,\, 2\nu_{3} \, - \, 2 \,\,\, , \quad \quad
\limsup_{\omega_{B} \tau \rightarrow \infty} \,\, 
{\log \sigma^{yy} (B) \over \log \omega_{B} \tau} 
\,\,\, = \,\,\, 2\nu_{2} \, - \, 2 $$
in the limit of strong magnetic fields. This trend, however, represents 
a very important general characteristic of the behavior of the quantities 
$\, \sigma^{xx} (B) \, $ and $\, \sigma^{yy} (B) \, $, since their exact 
dependence satisfies actually a number of important restrictions.

  We note here, in addition, that the presence of the 
Zorich - Kontsevich - Forni indices for chaotic trajectories of Dynnikov type 
also allows us to write the following relation
$$\limsup_{\omega_{B} \tau \rightarrow \infty} \,\, 
{\log | \Delta s^{xy} (B) | \over \log \omega_{B} \tau} 
\,\,\,\,\, \leq \,\,\,\,\, 
\nu_{2} \, + \, \nu_{3} \, - \, 2 $$
for the contribution of such trajectories to the off-diagonal term of 
the symmetric part of the conductivity tensor in the plane orthogonal to 
$\, {\bf B} $, which can also serve as an estimate for the overall trend 
of the decrease of this quantity (see \cite{TrMian}).

\vspace{1mm}

\begin{figure}[t]
\begin{center}
\includegraphics[width=\linewidth]{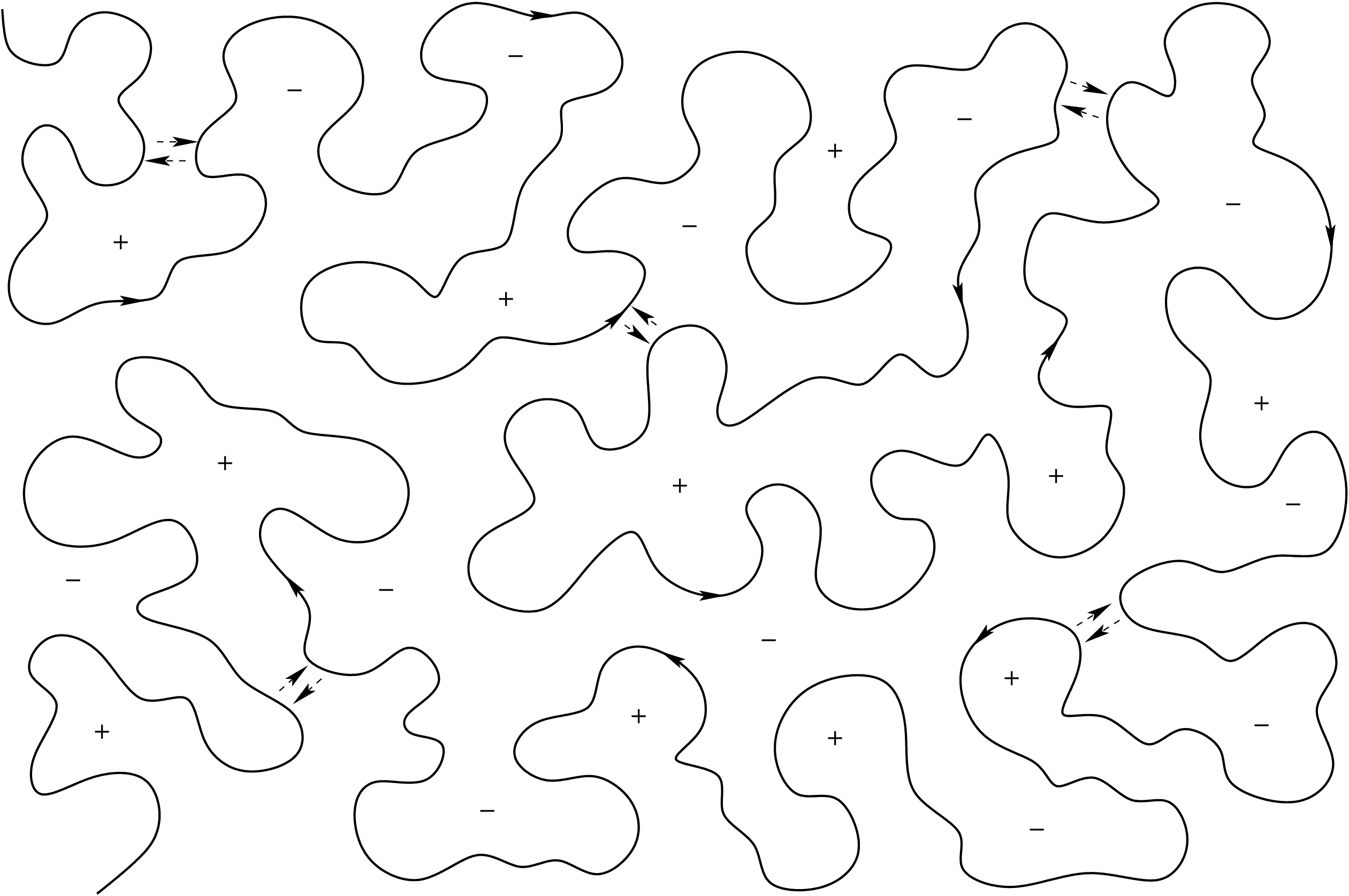} 
\end{center}
\caption{The phenomenon of magnetic breakdown on the Dynnikov chaotic 
trajectory in strong magnetic fields.}
\label{Proboi}
\end{figure}

 We would now like to note that all the above relations have been obtained 
within the framework of the kinetic theory on the basis of a purely 
quasiclassical analysis of the evolution of electron states in crystalline 
conductors. At the same time, as is well known, electron transport phenomena 
in sufficiently strong magnetic fields also have observable quantum 
corrections caused by quantum phenomena in electron systems 
(see e.g. \cite{Abrikosov,Kittel,etm,Ziman}). Here we would like to note that 
for Dynnikov chaotic trajectories (unlike most other trajectories of other 
types) the most significant of the quantum effects is the phenomenon 
of magnetic breakdown that arises in sufficiently strong magnetic fields. 
The phenomenon of (intraband) magnetic breakdown here is closely related 
actually to the presence of saddle singular points inside carriers of chaotic 
trajectories and consists in the possibility of jumps from one section 
of the trajectory to another (at a fixed $\, p_{z} $) for a sufficiently 
close approach of such sections to each other (Fig. \ref{Proboi}).

 The jump probability for two given sections increases with increasing 
of $\, B \, $ and tends to $\, 1/2 \, $ in the limit
$\, B \rightarrow \infty \, $. It is not difficult to see here that 
the mean time $\, \tau_{(1)}(B) \, $ of the electron motion between such 
jumps is a sufficiently rapidly decreasing function of $\, B \, $.
The effect of magnetic breakdown becomes significant when $\, \tau_{(1)}(B) \, $ 
turns out to be of order of $\, \tau \, $, and the problem represents one 
of the possible models of quantum chaos in the limit
$\, \tau_{(1)}(B) \, \ll \, \tau \, $. In the general case, the main effect 
of magnetic breakdown on transport phenomena in magnetic fields can be expressed 
by introducing the effective mean free time $\, \tau_{eff}(B) \, $, determined 
by the formula
$$\tau^{-1}_{eff} (B) \,\,\, = \,\,\, 
\tau^{-1} \, + \, \tau^{-1}_{(1)} (B) $$

  As already noted, the phenomenon of magnetic breakdown represents 
a purely quantum phenomenon in the electron system. For a more detailed 
description of the quantum picture arising in the presence of chaotic 
trajectories on the Fermi surface, it is first of all necessary to consider 
the spectrum of the one-electron states near the Fermi energy in this 
situation. Here it is most convenient to start the analysis of the 
one-electron spectrum from a purely quasiclassical spectrum, i.e. from 
the spectrum, described by the ``quantization'' of quasiclassical orbits 
in the $\, {\bf p} $ - space. As we told already, the Dynnikov chaotic 
trajectories exist only on one energy level $\, \epsilon_{0} \, $  
(see \cite{dynn1,dynn2,dynn3}), which should be close enough to the Fermi 
energy for the possibility of experimental observation of the regimes 
described above. At the energy levels $\, \epsilon < \epsilon_{0} \, $ 
all such trajectories break up into long closed trajectories of the 
electron type, and at the levels $\, \epsilon > \epsilon_{0} \, $ - into 
closed trajectories of the hole type. It can thus be seen that in the purely 
quasiclassical approximation (implying also the limit 
$\, \tau \rightarrow \infty $), we should expect the appearance of delocalized 
electron states only at the level $\, \epsilon = \epsilon_{0} \, $, while 
electron states near the level $\, \epsilon_{0} \, $ must be localized and 
determined by the quantization on the long closed trajectories.

 In accordance with the rules of the quasiclassical quantization 
(see \cite{Abrikosov,Kittel,etm,Ziman}, the closed trajectories of the 
system (\ref{MFSyst}) for each $\, p_{z} \, $ should be selected according 
to the rule
$$S (\epsilon, p_{z}) \,\,\, = \,\,\,
{2 \pi e \hbar B \over c} \, \left( n + {1 \over 2} \right)
\,\,\, ,  \quad  n \gg 1 \,\,\, , $$
where $\, S (\epsilon, p_{z}) \, $ represents the area bounded by a closed 
trajectory in the plane orthogonal to $\, {\bf B} \, $ in the 
${\bf p}$ - space.

\begin{figure}[t]
\begin{center}
\includegraphics[width=\linewidth]{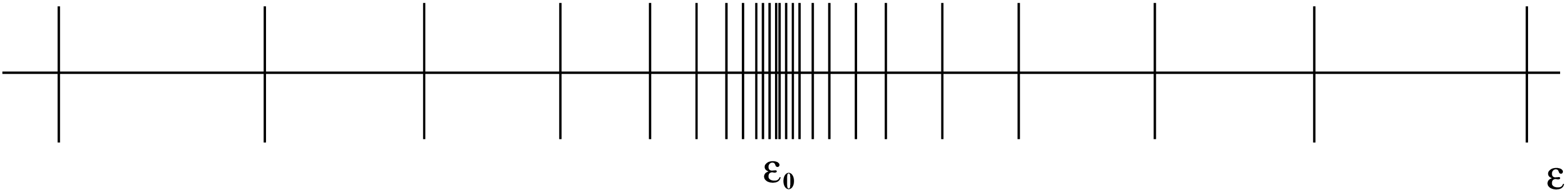} 
\end{center}
\caption{Quasiclassical electron levels near a special energy level 
carrying chaotic trajectories of the system (\ref{MFSyst}).}
\label{Levels}
\end{figure}

 It is easy to see that the areas of long closed trajectories 
$\, S (\epsilon, p_{z}) \, $ tend to infinity as 
$\, \epsilon \rightarrow \epsilon_{0} \, $, which leads to a rapid decrease 
in the distance between electron levels in the same limit (Fig. \ref{Levels}). 
Such a behavior of the arising quasiclassical levels makes it possible to easily 
distinguish their contribution to a number of quantum corrections on the background 
of the contribution of electron levels arising on ``short'' closed trajectories of 
the system (\ref{MFSyst}), the distance between which considerably exceeds 
the distance between levels arising on long closed trajectories of (\ref{MFSyst}).
Let us note here also that in studying the phenomena associated with the 
quantization of electron levels, usually the levels associated with 
the extremal trajectories on the Fermi surface, satisfying the condition
$\, \partial S / \partial p_{z} \, = \, 0 \,\,\, , $ 
are revealed.

 In the absence of the magnetic breakdown, the level structure shown 
at Fig. \ref{Levels}, is destroyed by scattering on impurities under the 
condition $\, T \, \geq \, \tau \, $ near the level $\, \epsilon_{0} \, $.
Scattering on impurities is associated with the electron motion in the 
coordinate space and ``mixes'' electron levels with different 
$\, p_{z} \, $, making the problem of determining of the new level 
structure essentially three-dimensional. It can be seen, however, that 
for a sufficiently large value of the magnetic field, we can have a much 
wider region near the value $\, \epsilon_{0} \, $, where the conditions
$\, T \, \geq \, \tau_{(1)}(B) \, $ and $\, \tau_{(1)}(B) \, \ll \, \tau \, $
are fulfilled. In the corresponding region, the change in the structure 
of the electron levels will be completely determined by the phenomenon 
of the magnetic breakdown and will be described by the behavior of the 
levels of two-dimensional systems for fixed values of $\, p_{z} \, $.
Systems of this type can be considered as one of the important models 
of quantum chaos, where a two-dimensional electron system has special 
quasiperiodic properties. The study of the structure of electron levels, 
as well as the transport properties of such systems, represents an important 
problem here both from the point of view of the mathematical theory 
of quantum chaos and from the point of view of the condensed matter physics.

\begin{figure}[t]
\begin{center}
\includegraphics[width=\linewidth]{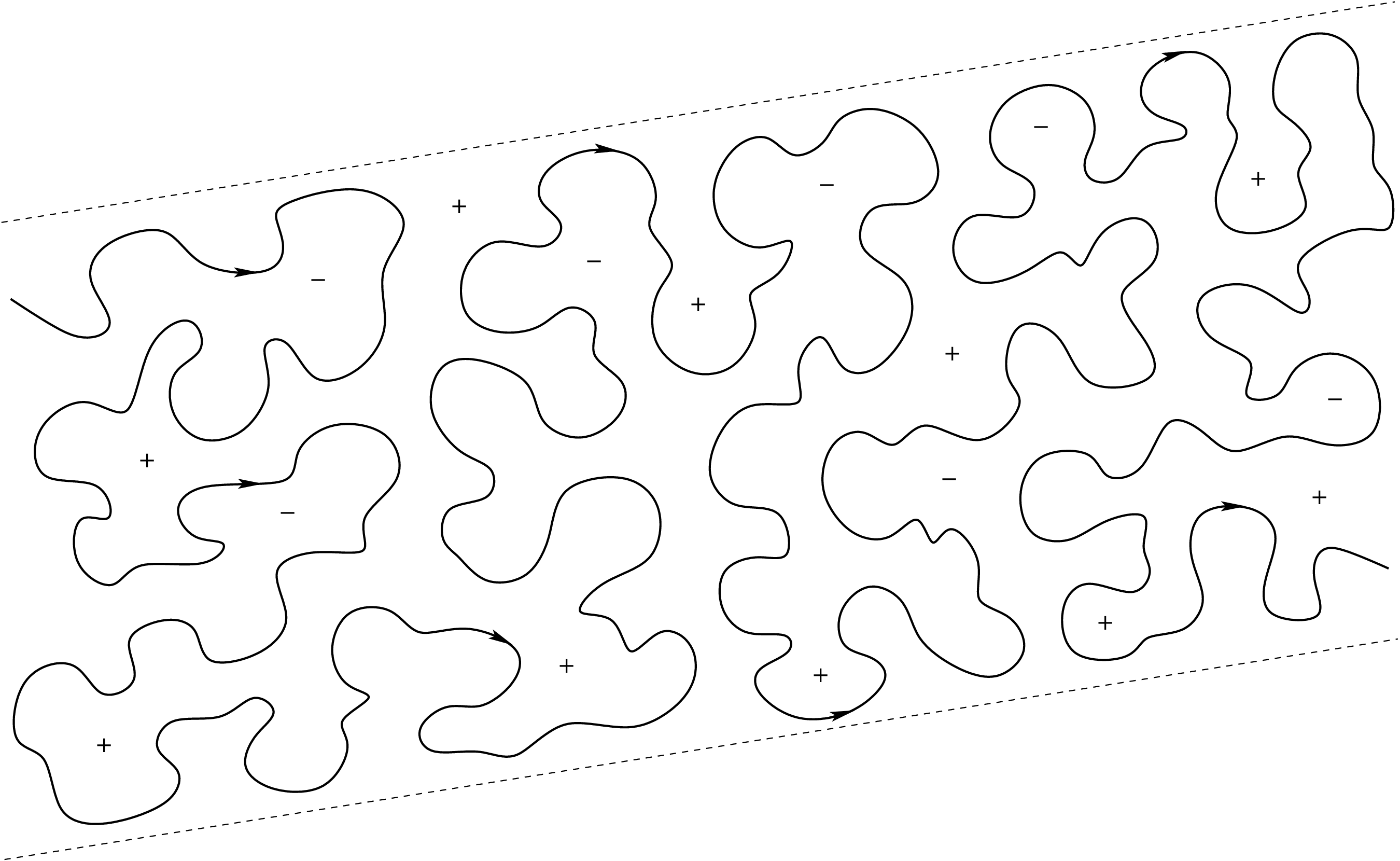} 
\end{center}
\caption{A stable open trajectory in the Zone $\, \Omega_{\alpha} \, $ 
of small sizes.}
\label{WideStr}
\end{figure}

 As already noted, all the above regimes are observed for special directions 
of  $\, {\bf B} \, $, corresponding to the appearance of chaotic trajectories 
of Dynnikov type on the Fermi surface. It must be said that for many of 
the real substances, such directions may in fact not be present on the 
angular diagram. In the general case, as was shown by I.A. Dynnikov 
(see \cite{dynn2,dynn3}), the Lebesgue measure of the corresponding 
directions on the angular diagram for generic Fermi surface is equal
to zero. According to the conjecture of S.P. Novikov 
(\cite{BullBrazMathSoc,JournStatPhys}), the fractal dimension of the set 
of such directions on the angular diagram for generic Fermi surface is 
strictly less than 1 (but may be larger for special Fermi surfaces).
Nevertheless, for substances with a rather complex Fermi surface, it is 
quite possible to expect an experimental observation of the described 
regimes for specially selected directions of $\, {\bf B} \, $.
In particular, the appearance of chaotic trajectories at levels arbitrarily 
close to the Fermi level should always be observed on the angular diagrams 
of type B described in the previous chapter. It should also be noted that 
stable open trajectories, corresponding to sufficiently small Stability Zones 
$\, \Omega_{\alpha} \, $, can also have a rather complex shape 
(Fig. \ref{WideStr}) and have the features of both regular and chaotic 
behavior depending on the interval of the values of the parameter 
$\, \omega_{B} \tau \, $.

\vspace{1mm}

 In conclusion of this chapter we consider briefly one more application 
of the Novikov problem to two-dimensional electron systems, which is actively 
studied in modern experiments. Namely, we consider the problem of 
a two-dimensional electron gas with a high mobility of carriers placed 
in the artificially created potential $\, V ({\bf r}) \, $. We note at once 
that at the present time there are many different methods of creating 
potentials of this type, many of them, in fact, are based on the use of 
superposition of certain periodic structures formed in the plane of the 
electron system. One of the most common methods, in particular, is the use 
of superposition of (one-dimensional) interference patterns of laser 
radiation, which causes the polarization of atoms forming the sample 
(see e.g. \cite{WKPW}).

\begin{figure}[t]
\begin{tabular}{ll}
\includegraphics[width=0.45\linewidth]{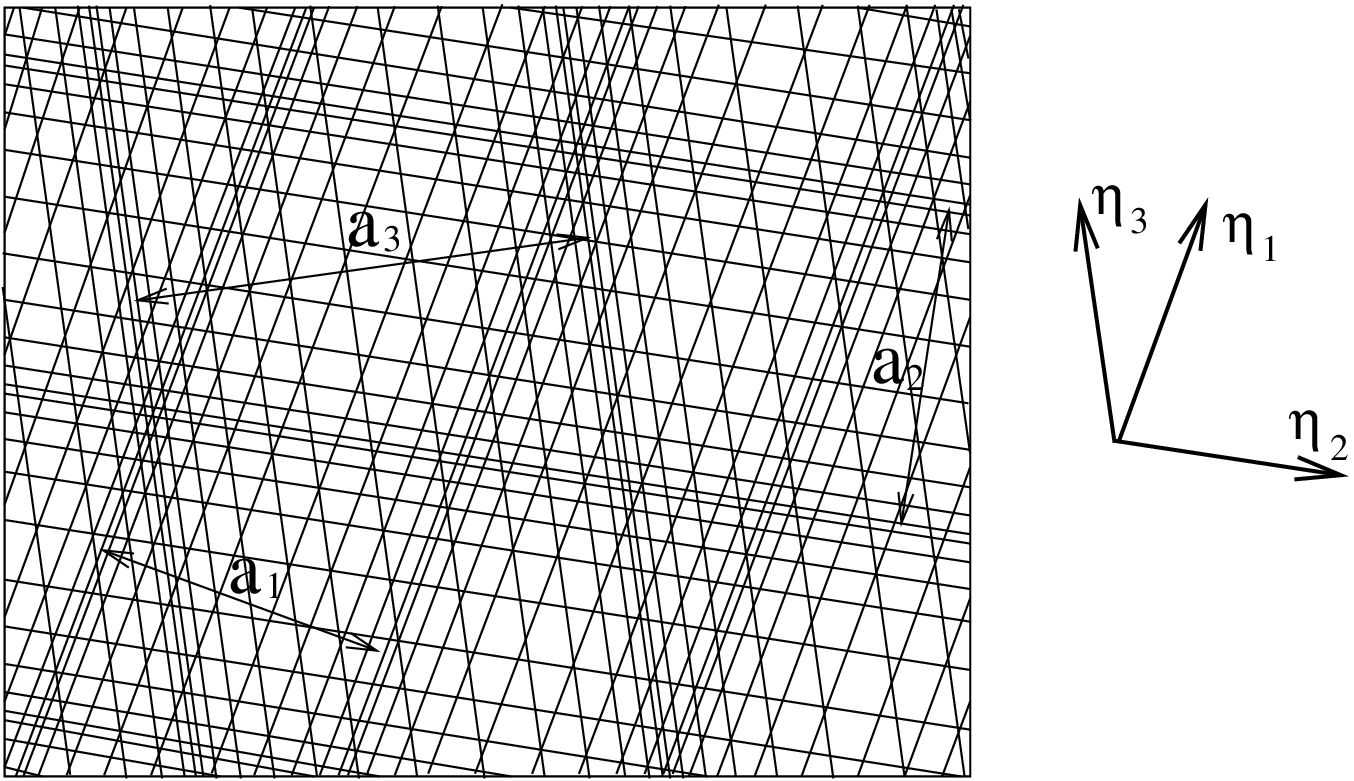}  &
\hspace{5mm}
\includegraphics[width=0.45\linewidth]{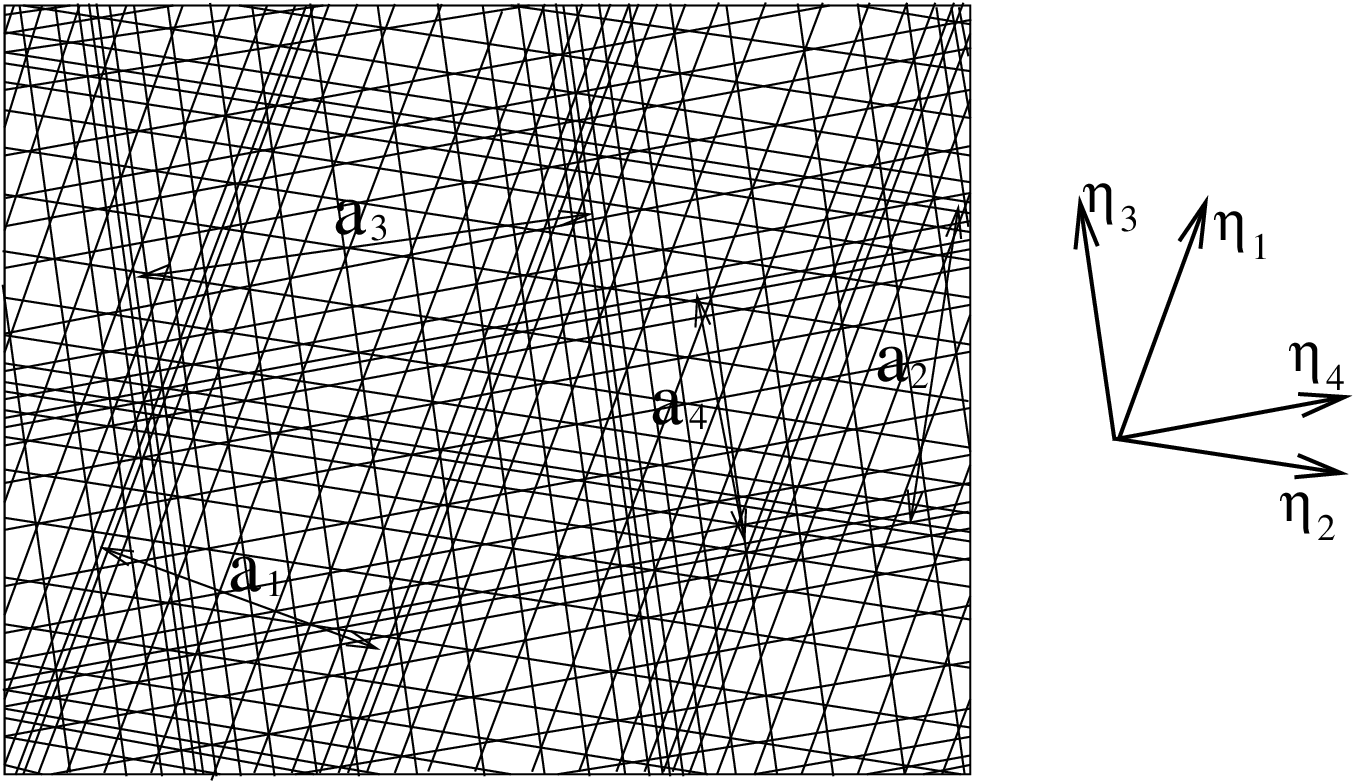} 
\end{tabular}
\caption{Quasiperiodic potentials with three and four quasiperiods, 
obtained by superposition of three and four one-dimensional potentials 
with different directions and period values, respectively.}
\label{QuasiPer}
\end{figure}

\begin{figure}[t]
\begin{center}
\includegraphics[width=\linewidth]{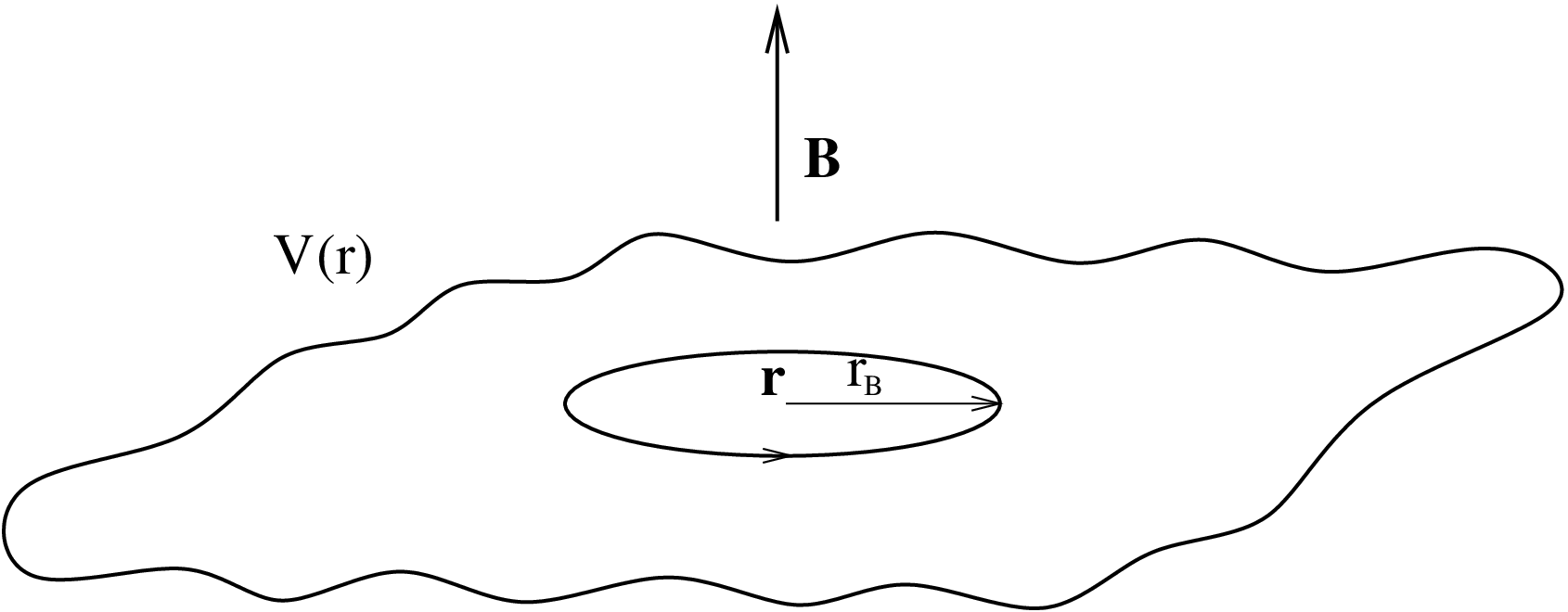} 
\end{center}
\caption{The cyclotron orbit in the presence of a quasiperiodic potential 
$\, V({\bf r}) \, $.}
\label{cyclorb}
\end{figure}

 The Novikov problem in these systems arises in the description of 
electron transport phenomena in the presence of a sufficiently strong 
magnetic field $\, {\bf B} \, $, orthogonal to the sample plane. 
The quasiclassical consideration of the electron motion 
(see e.g. \cite{Fertig,Beenakker}) allows one to use here the approximation 
of quasiclassical cyclotron orbits whose centers drift in the presence 
of the potential $\, V ({\bf r}) \, $ (Fig. \ref{cyclorb}). As can be 
shown, the drift of cyclotron orbit centers occurs along the level lines of 
the potential $\, V ({\bf r}) \, $ averaged over the corresponding cyclotron 
orbits. As in the case of normal metals, the main role here is played by 
electrons with energy close to the Fermi energy, so that for the electron 
motion one can also introduce a fixed cyclotron radius $\, r_{B} \, $ 
corresponding to the given problem. It is not difficult to see that 
the averaged potential $\, \bar{V}_{B} ({\bf r}) \, $ has then the same 
quasiperiodic properties as the potential $\, V ({\bf r}) \, $, and the 
description of the motion of orbits along the level lines of such a potential 
represents the Novikov problem with the corresponding number of quasiperiods.

 The great similarity between the two problems presented above makes 
it possible to transfer many results from the theory of the conductivity 
of normal metals to the theory of transport phenomena in the described 
two-dimensional systems (see e.g. \cite{JMathPhys}). We would like to note 
here that in the situation of two-dimensional systems, a much larger number 
of parameters of the system is controlled, which makes it possible to actually 
implement practically any of the interesting cases that arise in the 
investigation of the Novikov problem. In particular, two-dimensional electron 
systems can be a convenient experimental tool for studying chaotic regimes 
for specially selected parameters of such systems. It can also be noted here 
that by specifically choosing the parameters of the system, one can implement 
here both a purely quasiclassical case of chaotic behavior and a magnetic 
breakdown mode on chaotic trajectories. The latter circumstance is actually 
due to the fact that the semiclassical description here is also approximate, 
and here also jumps of cyclotron orbits between level lines of  
$\, \bar{V}_{B} ({\bf r}) \, $, close enough to each other, are possible
(see \cite{Fertig}). As can be shown, different behavior modes 
(in the limit $\, \tau \rightarrow \infty $) are determined here by external 
parameters of the system. In general, it can be noted that the described 
two-dimensional electron systems can represent a very convenient experimental 
base for studying both regular regimes in magnetotransport phenomena and the 
models of classical and quantum chaos described above.

 In conclusion, let us formulate a theorem that defines an important class 
of quasiperiodic functions on a plane with four quasiperiods whose level 
lines have properties analogous to the properties of stable open trajectories 
of the system (\ref{MFSyst}) (see \cite{NovKvazFunc,DynNov}). We give here, 
in fact, only the main corollaries of the results obtained in 
\cite{NovKvazFunc,DynNov}, a more detailed description of the resulting 
topological picture can be found in the papers \cite{NovKvazFunc,DynNov}.
It can be recalled here that a quasiperiodic function on a plane with four 
quasiperiods $\, f \, $ is defined as the restriction to the plane of some 
4-periodic function $\, F \, $ in four-dimensional space under a linear 
embedding $\, \mathbb{R}^{2} \subset \mathbb{R}^{4} \, $. 
According to \cite{NovKvazFunc,DynNov}, the following assertion can be stated:

\vspace{1mm}

1) All non-singular open level lines of the corresponding functions 
$\, f \, $ in $\, \mathbb{R}^{2} \, $ lie in straight strips of finite width, 
passing through them;

\vspace{1mm}

2) The mean direction of all non-singular open level lines 
of the functions $\, f \, $ is given by the 
intersection of the corresponding plane $\, \mathbb{R}^{2} \, $ with some 
integral three-dimensional plane $\, \Gamma_{\alpha} (\Pi) \, $ in the space 
$\, \mathbb{R}^{4} \, $, which is locally constant in the space
$\, G_{4,2} \, $. 

\vspace{1mm}

 It is not difficult to see that the statements formulated above correlate 
in a certain sense with the results obtained in the paper \cite{zorich1} for 
the case of quasiperiodic functions with three quasiperiods. It should also be 
noted that the proof of this theorem in the case of four quasiperiods requires, 
in fact, much greater effort.

\end{document}